\documentclass{aa}  

\usepackage{graphicx}
\usepackage{txfonts}
\usepackage{subfigure}
\usepackage{hyperref}
\begin{document}

   \title{Detection prospects of very and ultra high-energy gamma rays from extended sources with ASTRI, CTA, and LHAASO}
    \titlerunning{ASTRI, CTA, and LHAASO detection prospects}
    
   \author{S. Celli
          \inst{1,2,3}
          \and
          G. Peron
          \inst{4,5}
          }

   \institute{Sapienza Universit\`a di Roma, Dipartimento di Fisica,
                 P.le Aldo Moro 2, 00185, Roma, Italy
         \and
                 Istituto Nazionale di Fisica Nucleare, Sezione di Roma,
                 P.le Aldo Moro 2, 00185, Roma, Italy
        \and
                        Istituto Nazionale di Astrofisica,  Osservatorio Astronomico di Roma,
                Monte Porzio Catone, Roma, Italy
         \and
                Istituto Nazionale di Astrofisica, Osservatorio Astrofisico di Arcetri,
                L.go E. Fermi 5, Firenze, Italy 
     \and Université Paris Cité, CNRS, Astroparticule et Cosmologie, F-75013 Paris, France \\
                \email{silvia.celli@uniroma1.it, giada.peron@inaf.it}
     }


 
  \abstract
   {The recent discovery of several ultra high-energy gamma-ray emitters in our Galaxy represents a significant advancement towards the characterisation of its most powerful accelerators. Nonetheless, in order to unambiguously locate the regions where the highest energy particles are produced and understand the responsible physical mechanisms, detailed spectral and morphological studies are required, especially given that most of the observed sources were found to be significantly extended. }
   {In these regards, pointing observations with the next-generation Imaging Atmospheric Cherenkov Telescopes, such as the Cherenkov Telescope Array (CTA) Observatory and the ASTRI Mini-Array (ASTRI), are expected to provide significant improvements. Here we aim to identify the most promising sources to target in future observations.}
   {For this purpose, we performed a comparative analysis of the expected performance of ASTRI and CTA, computing their differential sensitivities towards extended sources, and further explored their capabilities with respect to specific case studies, including follow-ups of existing gamma-ray source catalogues.}
   {We find that almost all of the sources thus far detected by LHAASO-WCDA and in the H.E.S.S. Galactic Plane Survey will be in the reach of ASTRI and CTA with about 300 and 50~hours of exposure, respectively. For the highest energy emitters detected by LHAASO-KM2A, in turn, we provide a list of the most promising objects that would require further investigation. We additionally examined specific classes of sources in order to identify potentially detectable gamma-ray emitters, such as passive molecular clouds (i.e. illuminated by the cosmic-ray sea) and pulsars surrounded by a halo of runaway particles.}
    {} 

   \keywords{Radiation mechanisms: non-thermal -- 
            Catalogs --
            ISM: cosmic rays --
            ISM: clouds --
            Telescopes --
            Astroparticle physics
    }

   \maketitle
%
\section{Introduction}
In recent years, gamma-ray astronomy has witnessed a significant experimental boost. Ground-based instruments as HAWC and Tibet-AS$\gamma$ have started to survey the sky at very-high energies (VHEs, i.e. 100~GeV $\lesssim E \lesssim$ 100~TeV) and ultra-high energies (UHEs, i.e. $E \gtrsim$ 100~TeV) \citep{hawc2017,tibetCrab}, unveiling the most powerful northern sources. In 2021, the Large High Altitude Air Shower Observatory (LHAASO) started operating in its full configuration: thanks to its unprecedentedly large effective area and efficient background rejection, LHAASO has enabled the discovery of a few tens of UHE  emitters \citep{Cao2023Catalog} in the northern hemisphere, just in the first year of data acquisition, improving on the sensitivity of the currently operating extensive atmospheric shower (EAS) detectors by about one order of magnitude.
The planned SWGO and ALPACA detectors \citep{swgo2019,ALPACA2022} aim to survey the southern hemisphere at similar energies, to complement the view of the UHE sky.

Meanwhile, significant developments are foreseen regarding Imaging Atmospheric Cherenkov Telescopes (IACTs), particularly with the ASTRI Mini-Array (Astrofisica con Specchi a Tecnologia Replicante Italiana, hereafter ASTRI) and the Cherenkov Telescope Array Observatory (CTAO, hereafter CTA). ASTRI is a nine-telescope array (one telescope is already operative) currently under construction in Tenerife (Spain) \citep{astri2022}. In its final configuration, which is expected to be completed in the next few years, ASTRI will be sensitive to energies up to 200 TeV with an improved sensitivity compared to the currently operating IACTs (MAGIC, VERITAS, and H.E.S.S.) above a few TeV \citep{2022icrc.confE.884L}. CTA will in turn consist of two arrays, one located in La Palma (Spain) and the other in Paranal (Chile), thus covering both the northern and the southern hemisphere. Once completed, CTA will reach a sensitivity one order of magnitude lower than the currently operating IACTs, spanning a broad energy interval ranging from a few tens of GeV to a few hundred TeV \citep{cta2019}. \\
The identification of the UHE sources observed by HAWC and LHAASO is still uncertain, partly due to the limited angular resolution of their detection technique. In fact, most of the observed sources have a large angular extension and are located in crowded regions of the Galactic Plane, thus increasing the chances for source confusion. 
The advantage of detecting photons with arrays of IACTs is above all their superior angular resolution ($\lesssim$~3~arcmin), a specificity of the stereoscopic technique \citep{aharonian2013}. This is a key feature for obtaining a clear source identification and to conduct precise morphological investigations capable of distinguishing the sites of particle acceleration from nearby regions (e.g. molecular clouds illuminated by escaped particles). Furthermore, the larger Field of View (FoV) of the next-generation IACTs (up to $\sim 10^{\circ}$ in diameter) will allow us to study and resolve large and crowded sources such as supernova remnants, pulsar wind halos, and superbubbles. \\
Nevertheless, IACT observations suffer from a degraded performance with respect to the ideal on-axis point-like case, due to both off-axis pointing and increased background in extended analyses, which also affects their minimum detectable source flux. As shown in \citet{ambrogi2018} for a previous CTA layout (the so-called 2Q-array), the extent of the worsening depends on energy, and it tends to disappear at the highest energies. Thus, a more accurate evaluation of the detection prospects and required exposure per source is necessary when dealing with extended sources. For this work, we revised those predictions by considering the most up-to-date CTA proposed layout, the so-called Alpha configuration\footnote{\url{https://www.cta-observatory.org/science/ctao-performance/}}, and we provide a new computation of the CTA differential sensitivity for extended sources. In addition, we compute the expected ASTRI and LHAASO differential sensitivities towards extended sources. Such investigations allow us to clearly define a sample of gamma-ray sources that are foreseen to be detectable by the next-generation IACTs, and therefore it will be crucial to outline an optimised observational plan. We start by summarising the detection techniques of these instruments in Sec.~\ref{sec:detectors}, and later describe their specific features in Sec.~\ref{sec:irf}. We then introduce the methods that we developed for computing  instrumental sensitivities towards extended source observations in Sec.~\ref{sec:sensitivity}, and discuss the obtained results. As an application of the derived minimum detectable fluxes, in Sec.~\ref{sec:science} we explain how we scanned existing gamma-ray source catalogues in the VHE and UHE bands to define promising candidates for observations, demanding tailored IACT observations aimed at obtaining improved morphological and spectral information. Particularly, we discuss perspectives for detecting PeVatrons, passive molecular clouds illuminated by cosmic rays, and TeV halos. We then summarise the most promising targets in Sec.~\ref{sec:discussion}, and we finally draw our conclusions in Sec.~\ref{sec:conclusion}. Four appendices complement the text: in Appendix~\ref{sec:appA}, we show the integral sensitivities of the considered instruments; in Appendix~\ref{sec:appB} we discuss IACT response functions to off-axis sources, in Appendix~\ref{sec:appC} we present IACT differential sensitivities in high-zenith angle observations; and finally in Appendix~\ref{sec:appD} we provide a comparison among our results and official point-like source sensitivities from ASTRI, CTA, and LHAASO, also exploring individual sub-array configurations of CTA.

\section{Detection methods of VHE gamma rays}
\label{sec:detectors}
The Earth atmosphere is opaque to gamma rays, consequently ground-based detection methods rely on the indirect measurements of the secondary particles that compose the cascades generated by primary photons. Two main detection techniques exist for measuring such emission: the IACT technique, based on the observation at night of the Cherenkov flashes induced by the superluminous shower particles traversing the atmosphere; and the EAS technique, aimed at collecting individual shower particles from ground. The discrimination of photon- from hadron-induced showers is a fundamental part of the measurements. In the following we briefly describe the different detection and background rejection methods implemented by the present and next-generation instruments, referring to the dedicated works from each Collaboration for further details.

\subsection{IACTs: CTA and ASTRI}

The future of IACTs foresees two new facilities. ASTRI at its completion will be an array of nine telescopes located at the Observatorio del Teide in Tenerife  (28$^\circ$17\textquoteright 60.00\textquotedblright N, 16$^\circ$ 30\textquoteright 20.99\textquotedblright W) at an altitude of approximately 2370 m. Each of the telescopes will have two mirrors in the Schwarzschild-Couder design, of 4.3 m and 1.8 m diameter respectively, whose size allows the array to best perform in the energy range around 10~TeV. The telescopes are equipped with a wide FoV camera, about $10^\circ$ in diameter, constituted by silicon photo multipliers \citep{Scudieri2022Astri, Catalano2018AstriCamera}.

CTA will be an observatory located at two sites: the northern hemisphere array, so-called CTA-North, whose designated area is in the Canary Island of La Palma (28$^\circ$45\textquoteright 43.7904\textquotedblright N, 17$^\circ$53\textquoteright 31.218\textquotedblright W), while the southern array, so-called CTA-South, will be in the Paranal plateau in Chile (4$^\circ$41\textquoteright 0.34\textquotedblright S, 70$^\circ$18\textquoteright 58.84\textquotedblright W). The final CTA layout (the \textit{Alpha} configuration) has been recently established: it consists of 4 Large-Size Telescopes (LSTs, one of which is already operating) and 9 Medium-Size Telescope (MSTs) for the northern array, versus 14 MSTs and 37 Small-Size Telescopes (SSTs) for the southern site. The diameters of the main reflecting surface of the telescopes will be of 23, 11.5, and 4.3 m respectively for the large, medium and small size type, while their FoV will be $>4.5^\circ$, $>7^\circ$, and $>8^\circ$ respectively \citep{cta2019}. The use of many SSTs, distributed on a large ($\sim$3 km$^{2}$) area, is expected to improve the detection at high (5-300 TeV) energies, while the employment of the LSTs will allow a better sensitivity to be reached at lower (20--150 GeV) energies. Further improvements with respect to the Alpha configuration are expected as a result of the CTA+ program\footnote{\url{https://pnrr.inaf.it/progetto-ctaplus/}}, that has been recently approved and funded for the procurement of two more LSTs and five SSTs specifically for the CTA-South site \citep{Antonelli:2023uig}. \\

\subsection{EAS arrays: LHAASO}
\label{subsec:eas}
LHAASO \citep{2021arXiv210103508L,2021ChPhC..45b5002A} is an EAS observatory built at 4400~m above sea level in the Sichuan province of China. It consists of three type of detectors designed for the study of Cosmic Rays (CRs) and gamma rays across a broad energy range from sub-TeV to beyond the PeV energy scale: the Water Cherenkov Detector Array (WCDA), the Square Kilometer Array (also called KM2A), and the Wide Field-of-view Cherenkov Telescope Array (WFCTA). The WCDA is a 78000~m$^2$ surface water Cherenkov pond, filled with purified water and equipped with upward facing photomultiplier tubes (PMTs). The KM2A is an array composed of 5195 Electromagnetic Detectors (EDs) and 1188 Muon Detectors (MDs), extending for 1.3~km$^2$ surface. EDs are plastic scintillators detecting the electromagnetic component of EASs, covered by a lead plate ($0.5$~cm thick) and a $1.5^{\prime\prime}$ PMT. MDs are in turn cylindric muon tanks, filled with pure water and buried under 2.5~m of soil, as to absorb the electromagnetic component and detect the muonic component of an EAS. These two instruments operate independently, with a core energy range  of $\sim 100$~GeV to $\sim 50$~TeV for WCDA, and between $\sim 10$~TeV and $\sim 10$~PeVs for KM2A. Finally, 12 wide field of view air Cherenkov telescopes compose the WFCTA, complementing the system in CR observations; given that these are not used in gamma-ray related studies, we do not discuss WFCTA further. 
The observatory high sensitivity, particularly at the highest energies probed by KM2A and so far unexplored, was essential to detect the emission of PeV photons from several Galactic accelerators \citep{lhaaso1}.

\section{Instrument response functions}
\label{sec:irf}
Comparing instruments relying on different detection techniques and operating in different energy ranges is only possible through the evaluation of differential sensitivities, which establish, independently for every instrument, the minimum detectable flux per energy bin \citep{aharonian1991}.
The differential sensitivity computation, presented in the following section, is based on certain parameters, namely the same energy binning, observation time and detection thresholds, as well as on individual Instrument Response Functions (IRFs), specifically angular and energy resolutions, effective area, and expected background rate. We here proceed to the description of the latter quantities, according tho the latest available official releases by each Collaboration.

\subsection{LHAASO IRFs}
\begin{figure*}
    \centering
    \subfigure[\label{fig:psf}]{\includegraphics[width=0.49\textwidth]{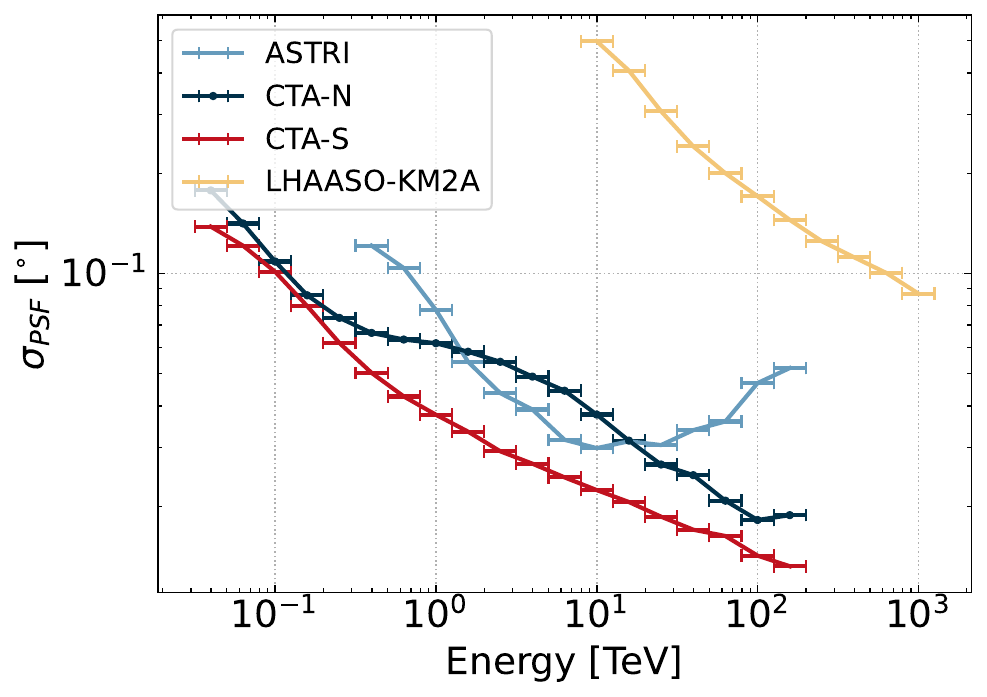}}
    \subfigure[]{\includegraphics[width=0.49\textwidth]{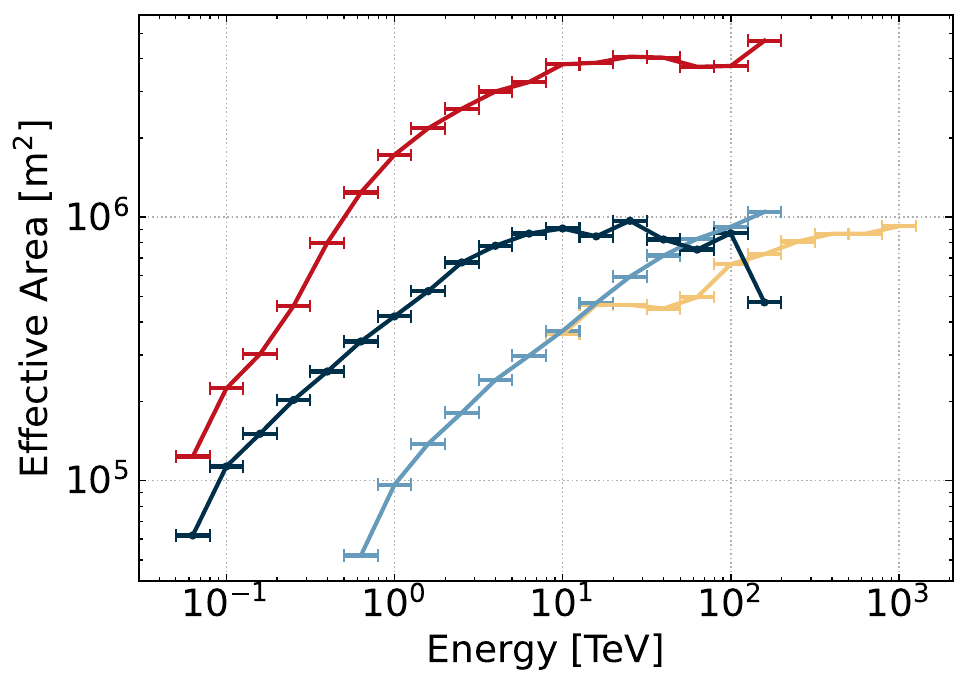}}
    \subfigure[\label{fig:bkg}]{\includegraphics[width=0.49\textwidth]{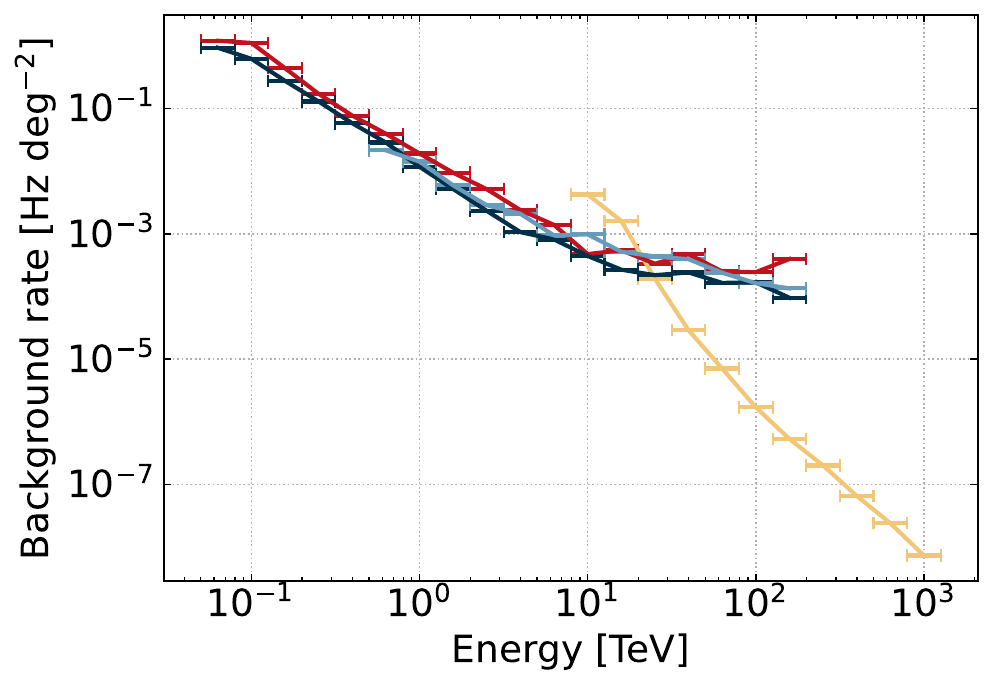}}
    \subfigure[]{\includegraphics[width=0.49\textwidth]{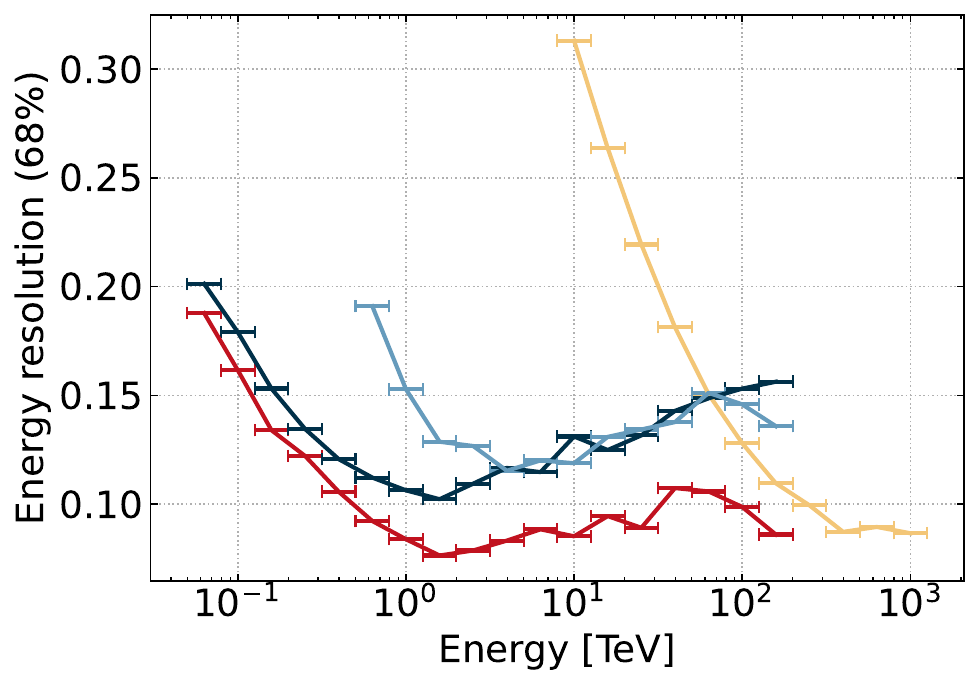}}
    \caption{Comparison between IRFs of LHAASO-KM2A \citep{km2aNew} and the on-axis ones of CTA (north and south) \citep{cta2022IRFs} and ASTRI \citep{astri2022IRFs}, as indicated in the legend, which applies to all panels. (a) 1-$\sigma$ width of the PSF approximated as a Gaussian distribution. (b) effective area, (c) background rate, and (d) energy resolution. The drop observable in the last point of the CTA-N effective area is an artefact of the simulations, resulting from a statistical fluctuation.}
    \label{fig:irfs}
\end{figure*}

We adopt the most recent response functions for LHAASO-KM2A, calibrated in the analysis of the Crab Nebula \citep{km2aNew}. We make use specifically of the IRFs relative to observations at zenith angle $\theta = 20^\circ$, to be consistent with those released by all of the experiments here investigated. Up-to-date performance of the WCDA detector is not yet available, thus we here limit our discussion to KM2A.
We consider the response functions as angular resolution, effective area to gamma rays, and gamma/hadron separation efficiency, all provided at analysis level, that is, including both cuts for hadron rejection as well as quality cuts selecting well reconstructed events. The IRFs released so far use event selection cuts optimised for point sources as the optimisation for extended sources requires additional care.  We expect analyses of extended sources to require more stringent cuts to reduce the background rate, which impacts the effective area. Hence, they require a dedicated investigation by the collaboration itself, that may be available in the next future. \\
While gamma-ray effective area, angular/energy resolution are provided in the IRFs, the expected background rate can be derived by convolving the all-particle CR flux (taken from \citet{gaisser}) with the effective area to hadronic showers at analysis level. The latter is obtained directly from the available photon effective area, which is first divided by the gamma-ray selection efficiency, and then multiplied for the CR rejection power as provided in \cite{km2aNew}. We finally integrate the resulting area with the all-particle CR flux and derive the quantity shown as a yellow line in Fig.~\ref{fig:bkg}. It is worth noticing that LHAASO-KM2A obtains its high rejection power thanks to the usage of muons for hadron identification: this is particularly important at the highest energies, where the increased abundance of muons facilitates the identification of the nature of primary particles.

\subsection{CTA and ASTRI IRFs}
\label{subsec:iact}
Fig.~\ref{fig:irfs} also shows the IACT IRFs for on-axis observations (while IRFs for off-axis observations at different offsets from the FoV centre are shown in Appendix~\ref{sec:appB}) for both ASTRI and the two CTA observatories. The latter can be retrieved in publicly available zenodo repositories \citep{astri2022IRFs,cta2022IRFs}: because these are optimised for point-like sources, tailored event selection will need to be assessed in the future specifically for extended sources, and might provide improvements with respect to those adopted here. Also
nominal pointing is assumed in the simulations, with all telescopes pointing to the same direction. \\
In IACT observations, hadron-initiated showers constitute a background for photon-induced cascades, hence they need to be rejected after the data acquisition \citep{2007APh....28...72M}. A first gamma/hadron separation is conducted based on the projected shape of the Cherenkov-light cone on the camera \citep{Hillas1985GammaHadron}: the high transverse momentum of secondary hadrons, in fact, induces a highly scattered distribution of hadronic showers at fixed energy of the primary. Nonetheless, an irreducible background of gamma-ray like hadron will survive the separation cuts. Such a residual hadronic emission is estimated at the level of the analysis, for example, by evaluating the background in a source-free region or by fitting a background template to the data in the region of interest \citep{Berge2007Background}.  
Recently, 3D likelihood statistical methods have been implemented also in VHE data analysis, showing improved results especially for low surface brightness sources or for objects in crowded regions \citep{Mohrmann2019Background,vovk2018}. The simplified approach used here is rather based on classical photon counting analysis, as discussed later in Sec.~\ref{sec:sensitivity}. We note that this choice will result into a more conservative estimation of source detectability.

A direct comparison among LHAASO, ASTRI, and CTA IRFs is now possible given that the same energy binning was set, namely 0.2 in logarithmic scale. The excellent angular resolution of IACTs, due to their pointing technique and camera, is visibly superior to the one of LHAASO-KM2A: in fact, the $1\sigma$ width of Point Spread Function (PSF) amounts to $\sigma_{\rm PSF}\lesssim 0.1^\circ$ for IACTs vs $\sigma_{\rm PSF} \gtrsim 0.1^\circ$ for LHAASO. With regards to the effective area, instead, for EAS arrays, it coincides with the geometrical area, namely the instrument surface exposed to the shower, since shower reconstruction relies on the number of particles collected on ground. In the case of LHAASO, a surface of 1~km$^2$ is covered by the KM2A. On the other hand, the effective area of IACTs is rather a convolution between the geometrical area and the Cherenkov light pool illuminated by the gamma-ray induced shower. The former depends on the specific layout of the experiment, namely by the relative distance among telescopes. For ASTRI, the foreseen layout covers an area of about $600 \times 300$~m$^2$ \citep{astri2022}, while CTA-North of about $600 \times 450$~m$^2$ and CTA-South of about $2000 \times 1800$~m$^2$ \citep{cta2019}. On the other hand, the Cherenkov pool of a primary photon interacting at a height of about 10~km spreads over a surface on ground of $3\times 10^4$~m$^2$ \citep{aharonian2013}: this value hence represents the effective area of an individual imaging telescope, which has to be rescaled to the number of telescopes when providing a description of the array. The effective areas shown in Fig.~\ref{fig:irfs} are obtained after the gamma/hadron separation cuts, while no selection relative to the reconstructed event direction was applied\footnote{\url{https://doi.org/10.5281/zenodo.5499840}}. \\

\section{Sensitivity studies towards extended sources}
\label{sec:sensitivity}
The calculation of differential sensitivities towards the observation of extended sources is here explained. This is motivated by the need to perform improved spectral studies of the newly detected UHE sources, as well as of the many unidentified sources which have emerged in VHE sky surveys. We highlight that, while the knowledge of the integral sensitivity is usually sufficient to evaluate the overall detection capabilities of an instrument, the differential approach enables a rigorous investigation of the performance towards spectral studies. In particular, the presence of cut-offs in the energy distribution of sources makes the differential approach the only reliable assessment of detection prospects beyond its pivot energy. On the other hand, the integral approach would tend to overestimate them, particularly in the case that a cut-off has not yet been clearly assessed. We hence proceed with a detailed discussion about differential sensitivity studies, while we refer the interested reader to Appendix~\ref{sec:appA} for the integral sensitivity computation.

Following the approach described in \citet{ambrogi2018}, we define the minimum detectable flux by imposing that the following conditions are satisfied in each energy bin of the computation \citep{2013APh....43..171B}, namely:
\begin{itemize}
\item[1)] a minimum number of signal events, $N_{\rm s}^{\rm min}=10$;
\item[2)] a minimum significance detection level, $\sigma_{\rm min}=N_{\rm s}/\sqrt{N_{\rm b}}=5$;
\item[3)] a minimum signal excess over the background, $N_{\rm s}^{\rm min}/N_{\rm b}=0.05$. 
\end{itemize}
The number of signal events  $N_{\rm s}$ expected per energy bin is obtained by folding the gamma-ray effective area with the Crab nebula spectrum\footnote{This is described in the TeV regime as a power-law function, $F_0 E^{-\alpha}_{1~\rm TeV}$, where $F_0= 2.83\times 10^{-11}$ (TeV  cm$^{2}$ s)$^{-1}$ is the flux normalisation at 1 TeV and $\alpha=2.62$ \citep{Aharonian2004Crab}.}. However, it should be noted that the final sensitivity computation is independent of the assumed source flux.
The number of background events $N_{\rm b}$ is computed from the background rate, assumed to be constant in time. While the background rate is directly provided in IACT IRFs, in the case of LHAASO we convolved the expected background flux with the effective area calculated for hadrons, as explained in the previous section. For the significance, $\sigma$, we here adopt the definition given by \citet{lima}, with $N_{\rm on}=N_{\rm s}+N_{\rm b}$ and $N_{\rm off}=N_{\rm b}/\alpha$, $\alpha$ being the ratio of the on-source to the off-source exposure; for our calculations, we assume $\alpha=1$.

\begin{figure*}
    \centering
    \subfigure[]{\includegraphics[width=0.49\textwidth]{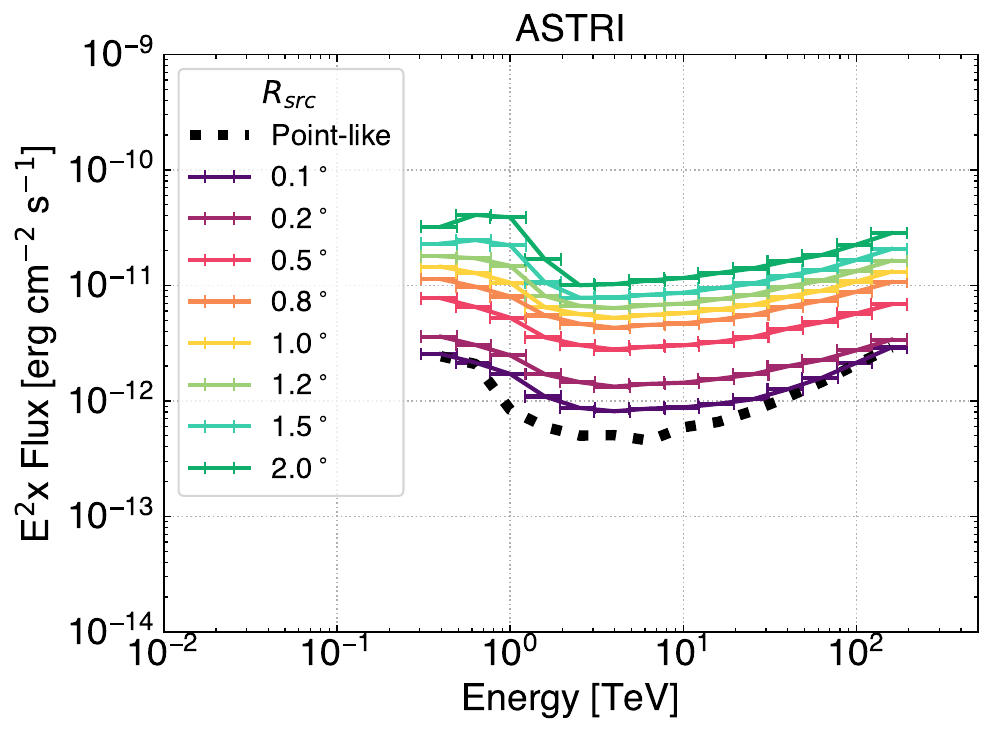}}
    \subfigure[]{\includegraphics[width=0.49\textwidth]{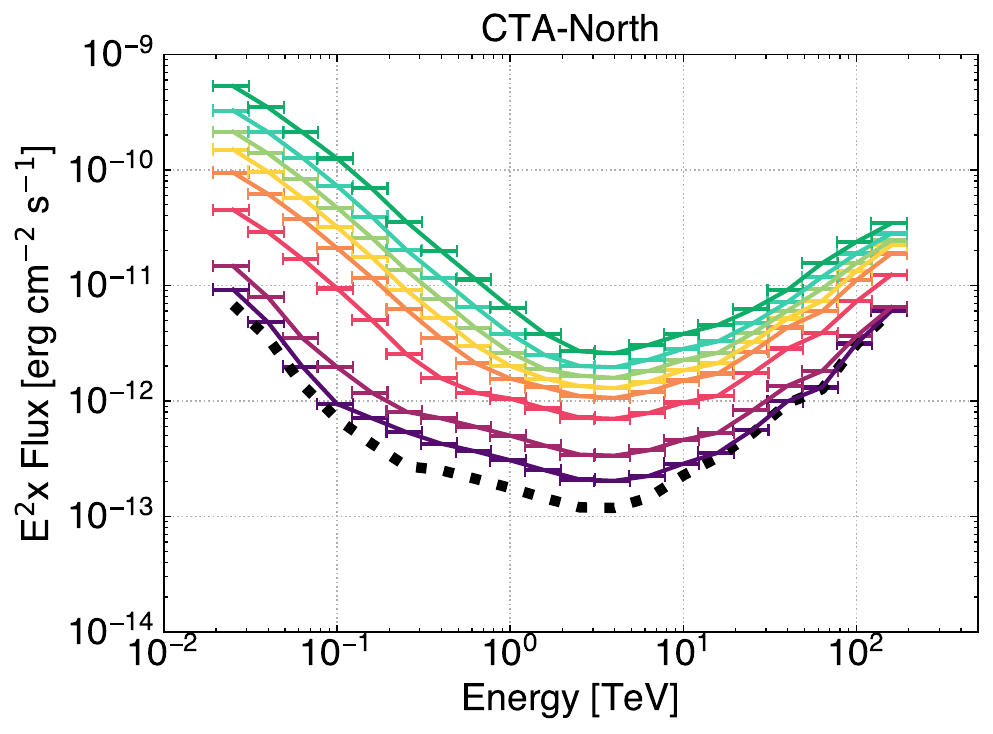}}
    \subfigure[]{\includegraphics[width=0.49\textwidth]{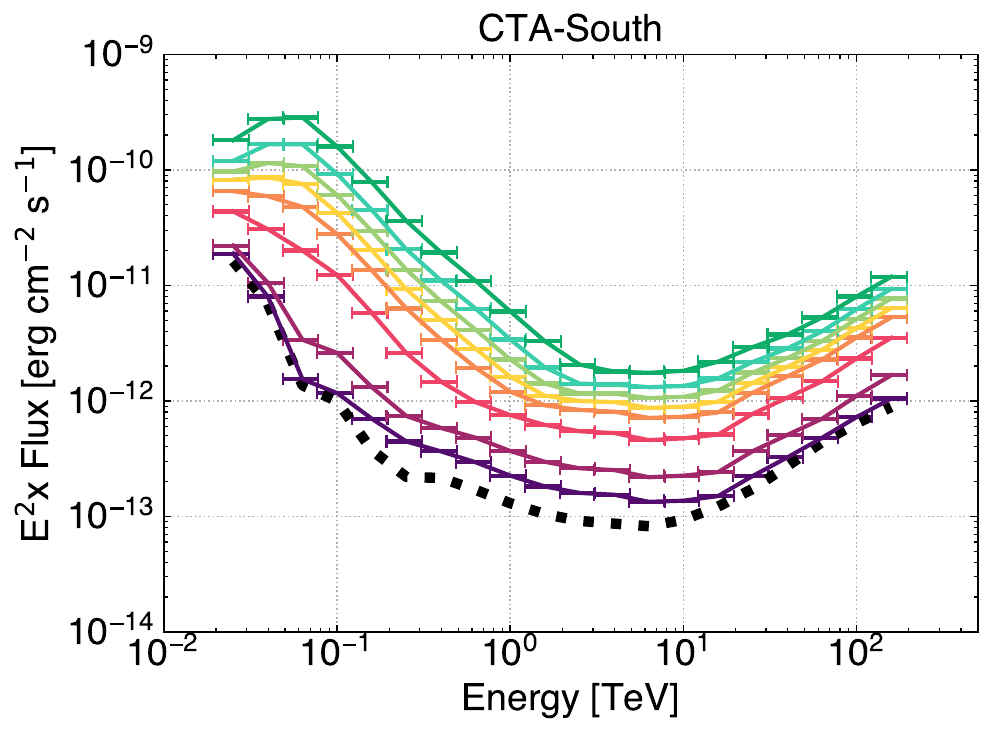}}
    \subfigure[\label{fig:sensL}]{\includegraphics[width=0.49\textwidth]{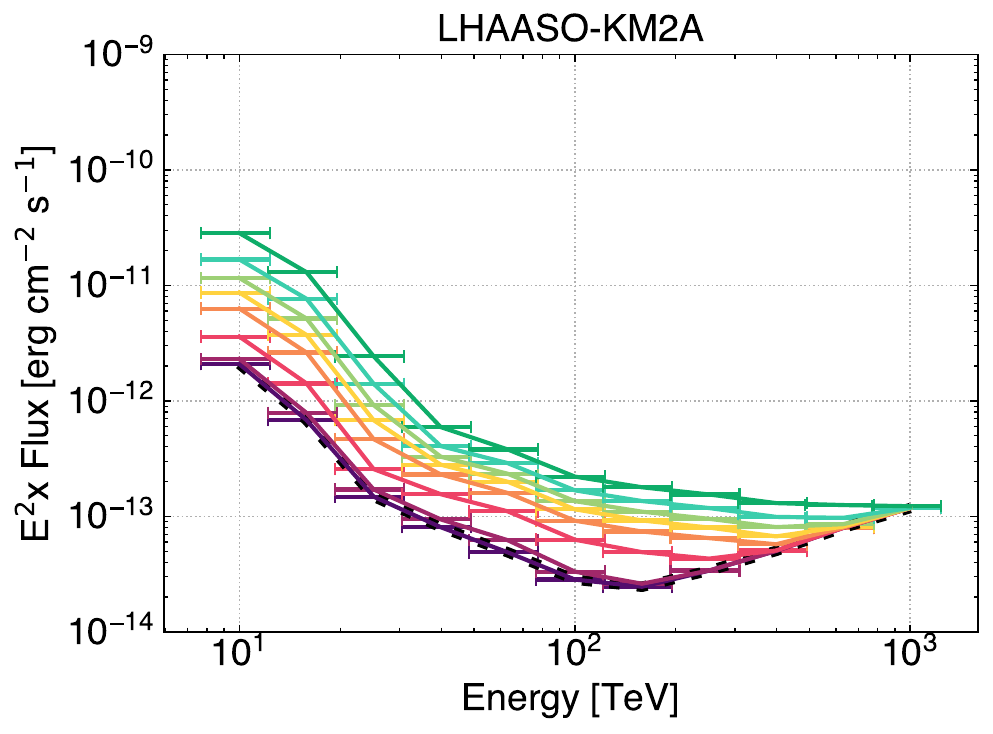}}
    \caption{Differential sensitivity towards the observation of extended sources by: (a) ASTRI, (b) CTA-N, (c) CTA-S, and (d) LHAASO. The exposure time adopted is 50 h for IACTs and 1 year for LHAASO. The source considered has a uniform disk-like morphology, whose angular radius is indicated in the legend, valid for all panels. The dotted line represents the point-like source sensitivity.}
    \label{fig:sens}
\end{figure*}

\begin{figure}
    \centering
    \includegraphics[width=1\linewidth]{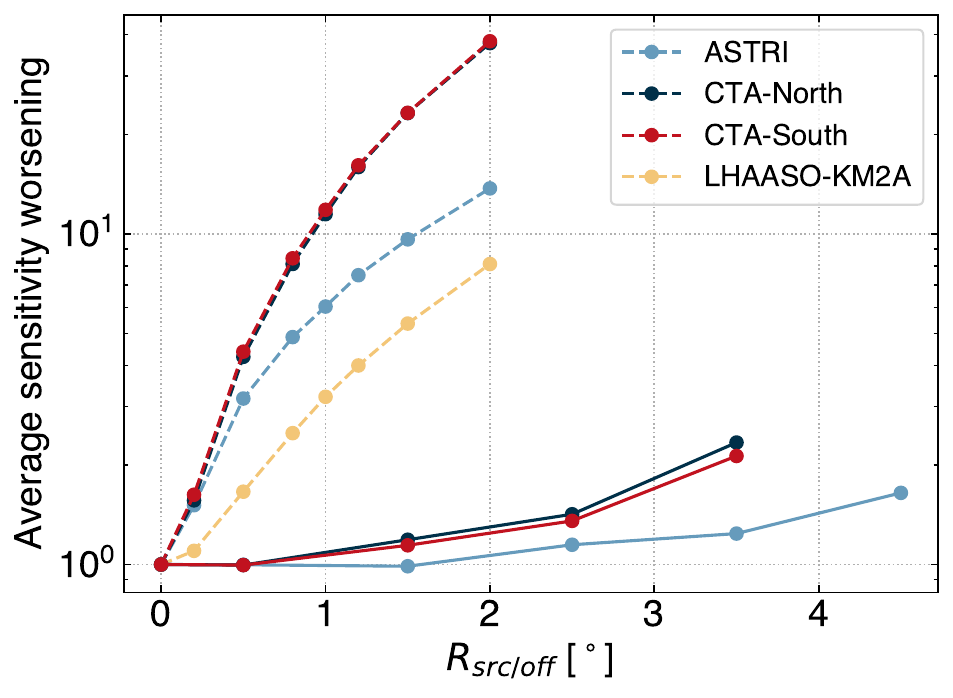}
    \caption{Worsening of the differential sensitivity 
    averaged for energies >1 TeV for IACTs and >10 TeV for LHAASO as a function of the source extension ($R=R_{\rm src}$, dashed lines) and of the offset ($R=R_{\rm off}$, solid lines). The worsening is expressed as the ratio between the calculated sensitivity and the on-axis point-like source sensitivity.}
    \label{fig:wors_sens}
\end{figure}

\begin{figure*}
    \centering
    \subfigure[]{\includegraphics[width=0.49\textwidth]{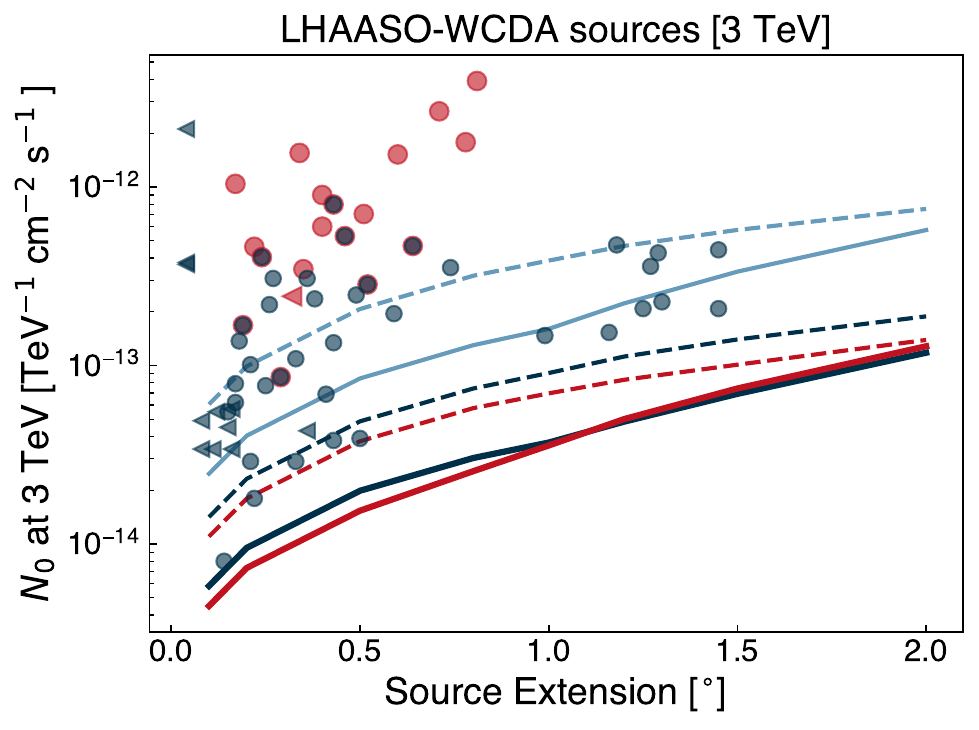}}
    \subfigure[\label{fig:lhaasoKM2A}]{\includegraphics[width=0.49\textwidth]{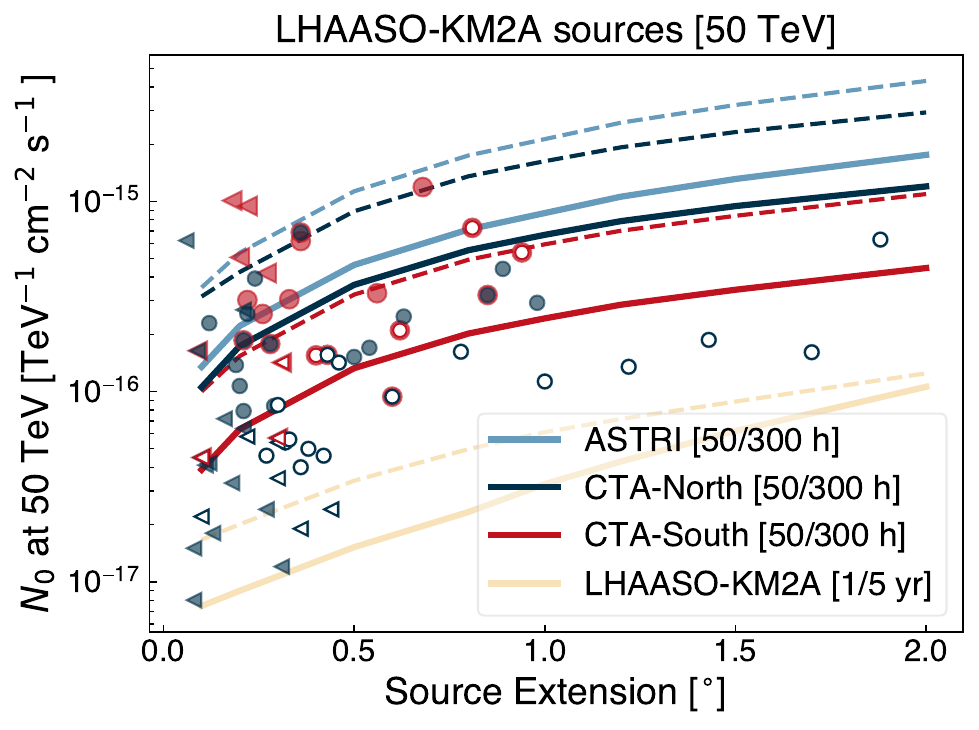}}
    \caption{Differential flux of LHAASO first catalogue sources as a function of their radius, measured (circles) or upper limits (triangles), as evaluated by (a) WCDA at 3 TeV and (b) KM2A at 50 TeV. The colour code refers to culmination at zenith angles below $30^\circ$ at each observatory sites (red for CTA-South, blue for CTA-North and ASTRI). Filled markers in the right panel correspond to UHE sources in LHAASO first source catalogue (with test statistics at 100~TeV ${\rm TS}_{100}>20$). Lines show the ASTRI and CTA differential sensitivities as a function of the source extension; dashed for shorter observation time and solid for the longer one, as indicated in the legend that applies to both panels.}
    \label{fig:visibilityLHAASOcat}
\end{figure*}

In each energy bin, the instrument sensitivity is limited by the one condition among the three listed above which dominates over the other two. As already discussed in \citep{ambrogi2018}, we find that source detection is determined by the signal over background ratio at low energies, by the  significance condition in the intermediate energy range (i.e. close to the sensitivity minimum), and by signal event counts at the highest energies where the statistics becomes low.
Since we require all the above criteria to be satisfied in each energy bin, a differential $5\sigma$ requirement corresponds to an actual higher significance in the energy bins where this is not the dominant condition (normally at the lowest and highest energies, respectively). 
Concerning the condition on the background uncertainty, for all detectors we assume a 1\% accuracy on the modelling (and hence subtraction) of the residual background and require a signal excess of at least five times this background systematic uncertainty, namely $ N_{\rm s}/N_{\rm b} \geq 0.05$, following the approach adopted by ASTRI and CTA \citep{cta2019}.  \\
Moreover, to account for the statistical fluctuations of the background, the number of background events is randomly extracted from a Poissonian distribution with mean value $N_{\rm b}$, and the result shown is averaged over 1000 realisations of the sensitivity estimation. Given the different observation strategy of the instruments, a benchmark observation time of 50~h is assumed for all IACTs while 1~year is considered for LHAASO, unless otherwise specified. Furthermore, IRFs at an average zenith angle of 20$^{\circ}$ are used for the calculation of the extended source sensitivities, given that public IRFs are available for all the observatories under investigation specifically at this value. The effects of IACT observations at different zenith angles are discussed in Appendix~\ref{sec:appC}. We consider here disk-like sources with eight different radial sizes, namely $R_{\rm src} = [0.1, 0.2, 0.5, 0.8, 1.0, 1.2, 1.5, 2.0]^\circ$; the choice of these specific sizes is mostly driven by an analogue computation in our previous publication, see \citep{ambrogi2018}. However, we limit the discussion to sources with radius smaller than $2.0^\circ$ , which is much smaller than the FoV of any considered-instrument. Otherwise, for source sizes comparable to the instrument FoV, the background estimation should account for additional systematic effects beyond the statistical noise considered in the presented calculations. The instrument PSF affects the effective radius in which the signal is spread ($R_{\rm ON}$), which we calculated as the convolution between the image of the source, described as a disk of radius $R_{\rm src}$, and the PSF, assumed to be a Gaussian function with $\sigma_{\rm PSF}$ its standard deviation. Note that $\sigma_{\rm PSF}$ depends on energy (see Fig.~\ref{fig:psf}), and consequently also $R_{\rm ON}$.
For IACTs, the IRFs further depend on the offset from the centre of the camera, as described in Appendix~\ref{sec:appB}.
In order to account for the variation of performance across the FoV, we construct for each IRF a data cube, namely defining the corresponding value at each energy (E) and position (x,y) in the FoV. The source flux is analogously binned, with an energy dependence given by the spectral assumption. We proceed then with the determination of $R_{\rm ON}$ by convolving the image with the PSF, considering the proper kernel $\sigma_{\rm PSF}(x,y,E)$ for each point of the grid. We then compute the number of signal and background counts expected within $R_{\rm ON}$ by considering a weighted average of the effective area inside the on-region. With respect to the approach previously implemented in \citet{ambrogi2018}, this method allows us to account for the degradation of IRFs across the FoV and also to extend the computation to off-axis sources, that is centred at given distance $R_{\rm off}$ from the centre of the FoV. We discuss this point further in Appendix~\ref{sec:appB}, where we also present differential sensitivities at different offsets in Fig.~\ref{fig:offset_sens}. \\
Our results indicate that the differential sensitivity strongly depends on the source extension, as it can be seen from Fig.~\ref{fig:sens}. The worsening due to the source extension for all IACTs is found to be about one order of magnitude for most of the investigated sizes, while it remains more contained for KM2A, as expected due to its larger angular resolution. We also note that LHAASO-KM2A achieves a truly background free detection regime at energies above 1~PeV, as demonstrated by the convergence of sensitivity lines computed for different source extensions (see Fig.~\ref{fig:sensL}). 
On the other hand, the degradation due to the offset is restrained to a factor of a few units for all IACTs, as shown in Fig.~\ref{fig:offset_sens}. A comparative evaluation of the impact of these two effects (source extension and offset) is provided in Fig.~\ref{fig:wors_sens} , where the sensitivities computed for  different extensions ($R_{\rm src}$) and at different offsets ($R_{\rm off}$), for IACTs, are compared to the sensitivity towards observations of a point source at the centre of the camera ($R_{\rm src} = R_{\rm off}=0^\circ$). With regards to the sensitivity worsening induced purely by source extension (on-axis), we note that IACTs are more affected than EAS because of the intrinsic sharper PSF of the former. On the other hand, with regards to the offset, it is interesting to note that ASTRI sensitivity is almost unaffected by the offset, mainly because of the small variation of its IRFs across the FoV, as it can be seen in Fig.~\ref{fig:astri_offset_IRFs}. This guarantees the possibility of studying multiple sources in the same pointing without losing much in performance. \\
To test our methods, we compared our calculations with the sensitivity for point-like sources calculated by the CTA and ASTRI collaboration, finding consistent results (see Appendix~\ref{sec:appD} and Fig.~\ref{fig:sens_ps}). A consistency check among the resulting LHAASO-KM2A sensitivity and the sources listed in its first catalogue can be performed, shown in Fig and \ref{fig:lhaasoKM2A} and further discussed later on.

\section{Science cases}
\label{sec:science}
IACT and EAS measurements are complementary in many aspects: the former allow us to carry out precise spectroscopic and morphological studies of sources thanks to their superior energy and angular resolutions, while the large duty cycle and field of view of EAS arrays enable the deepest exposure to sources. The combination of these features will permit some of the most relevant key science cases for the community to be investigated with
an unprecedented level of accuracy. Specifically, the following developments are expected: \\
i) The improved angular resolution of IACTs will enable us to precisely locate the regions where particle interactions take place, within or around the accelerator, thus revealing information about the nature of the interaction itself (e.g. the presence of correlated emission with dense gas regions hints towards hadronic collision origin). This will further allow us to understand sources remaining unidentified  in current VHE and UHE catalogues, as the first LHAASO catalogue \citep{Cao2023Catalog} or in the H.E.S.S. Galactic Plane Survey (HGPS), both of which are discussed in Sec.~\ref{sec:catalog}. \\
ii) The improved energy resolution will be a key feature to investigate the cut-off region of energy spectra, containing information about in-situ acceleration, propagation and radiation properties of the primary particles producing the gamma rays \citep{Celli2020Pevatron}. A careful analysis of the cut-off region is of paramount importance in the PeVatron search, particularly with regards to the hadronic PeVatrons responsible for the CR flux observed at Earth, as we discuss in Sec.~\ref{ss:pevatrons}. \\
iii) The large FoV of the next-generation IACTs will further allow us to better constrain the background around sources, facilitating the detection and the characterisation of low-surface brightness sources such as molecular clouds, discussed in Secs.~\ref{ss:mc}. Together with the improved angular resolution, this will permit particle propagation around their sources to be modelled better: for example, we expect the source size to change in energy in a leptonic scenario because of energy losses, as observed in gamma rays from the microquasar SS433, for example \citep{hessMagicSS443,hawcSS443, hessSS433}, while in hadronic scenarios the most energetic particles will travel loss-free to a further distance before interacting. An application of the propagation and escape concept certainly concerns Pulsar Wind Nebulae (PWNe) and their surrounding TeV halos \citep[for example]{giacinti2020}, as we discuss in Sec.~\ref{ss:pwn}. \\
As a general remark for the following discussion, we mention that we do not explicitly report on the coordinates of the sources under investigation, for which we invite the interested reader to look at the detection papers by the respective collaborations.

\begin{figure}
    \centering
    \includegraphics[width=1\linewidth]{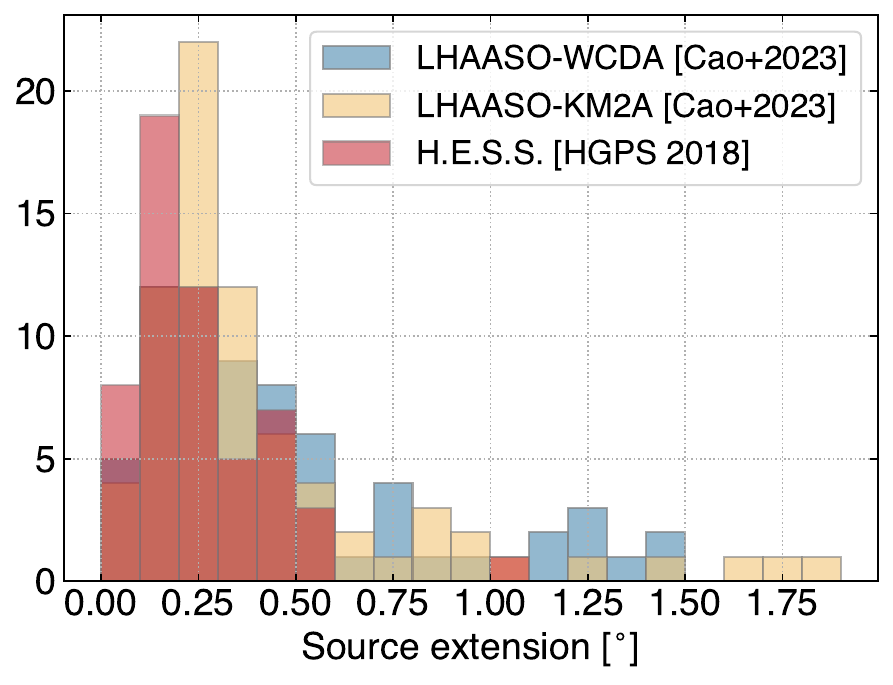}
    \caption{Distribution of the extension of the sources included in the HGPS \citep{hgps} and in the first LHAASO source catalogue \citep{Cao2023Catalog}.}
    \label{fig:extensions}
\end{figure}

\subsection{Catalogue follow-ups}
\label{sec:catalog}
Currently, both the HGPS \citep{hgps} and the first LHAASO source catalogue \citep{Cao2023Catalog} contain more than half unidentified sources, requiring follow-up studies with improved-performance instruments. 
We therefore proceed to the exploration of such source catalogues, in order to establish the most promising targets for the observatories under consideration. 

\subsubsection{The first LHAASO source catalogue}
\label{ssec:1stL}
After the announcement of the first 12 sources \citep{lhaaso1}, a new catalogue was recently published by the LHAASO Collaboration \citep{Cao2023Catalog}, reporting on the observation of 90 sources of both Galactic and extra-galactic nature. Among these, 32 represent new discoveries and 43 are significantly detected in the UHE domain\footnote{With a test statistics (TS$_{100}$) above 20 at 100 TeV, these sources are labelled with the letter $u$ in \citet{Cao2023Catalog}.}. The first LHAASO catalogue was obtained by analysing the WCDA and KM2A data-sets separately, and only 54 sources are significantly observed by both instruments. The differential spectra measured by the two instruments are fitted independently, each with a power-law shape and the best-fit parameters are reported in the catalogue.
Although a tentative association is proposed in \citet{Cao2023Catalog}, the large extensions (see Fig.~\ref{fig:extensions}) and superposition of sources, together with a poor understanding of the diffuse emission at such high energies, make the identification of the sources challenging. IACTs will at least partly compensate these issues, provided that a reasonable exposure will be guaranteed to detect source fluxes. 

Thanks to the methods developed in this work, we are able to assess which of these sources can be effectively followed up, also accounting for the proper sensitivity worsening in case of extended objects. 
We show in Fig.~\ref{fig:visibilityLHAASOcat} the differential fluxes of the catalogued LHAASO sources at 3~TeV and 50~TeV, provided respectively in the WCDA and KM2A catalogues, as a function of the measured source extension \citep{Cao2023Catalog}. We compared them to the sensitivity of future IACTs at the same energies and for different exposure times. The sources are further divided depending on their culmination zenith angles ($Z_{\rm min}= 
\mathrm{sign}(\Psi)(\Psi - \delta $), being $\Psi$ the geographical latitude of the observatory, and $\delta$ the declination of the source), where red and blue points in the figure represent sources that culminate at a zenith below 30$^\circ$ at the latitude of CTA-South and CTA-North (which is almost at the same latitude of ASTRI), respectively. \\
With regards to WCDA sources, we find that almost all of the (so-defined) southern sources will be accessible to CTA-South with 50~hr exposure, in line with the expectations. In fact, because the southern sources fall at a higher zenith angle at the LHAASO site, a selection bias towards the brighter sources might be in place. In other words, because the LHAASO exposure is suppressed for sources observed at high-zenith angles \citep{LHAASOcrab}, most likely only the brightest UHE sources in the southern hemisphere are included so far in the LHAASO sample. Many of the northern hemisphere sources will be seen by CTA-North and ASTRI with 50~hr exposure, and almost all with a longer exposure around 300~hr. 
With regards to KM2A sources, it appears that the extension of the sources might prevent in certain cases a clear detection with the considered observational setup. In particular, while many sources will be observed by CTA-South with 50~hr exposure, some will require a longer observation time, possibly up to $\sim 300$~h. Alongside, many of the detected KM2A-UHE sources will be detectable, allowing for a better characterisation of these accelerators. Deep (300~hr) observations with ASTRI and CTA-North will also achieve a good-enough sensitivity to allow the follow-up of a handful of these interesting sources already in the next few years.

\begin{figure*}
    \centering
    \subfigure[\label{fig:hessA}]{\includegraphics[width=0.49\textwidth]{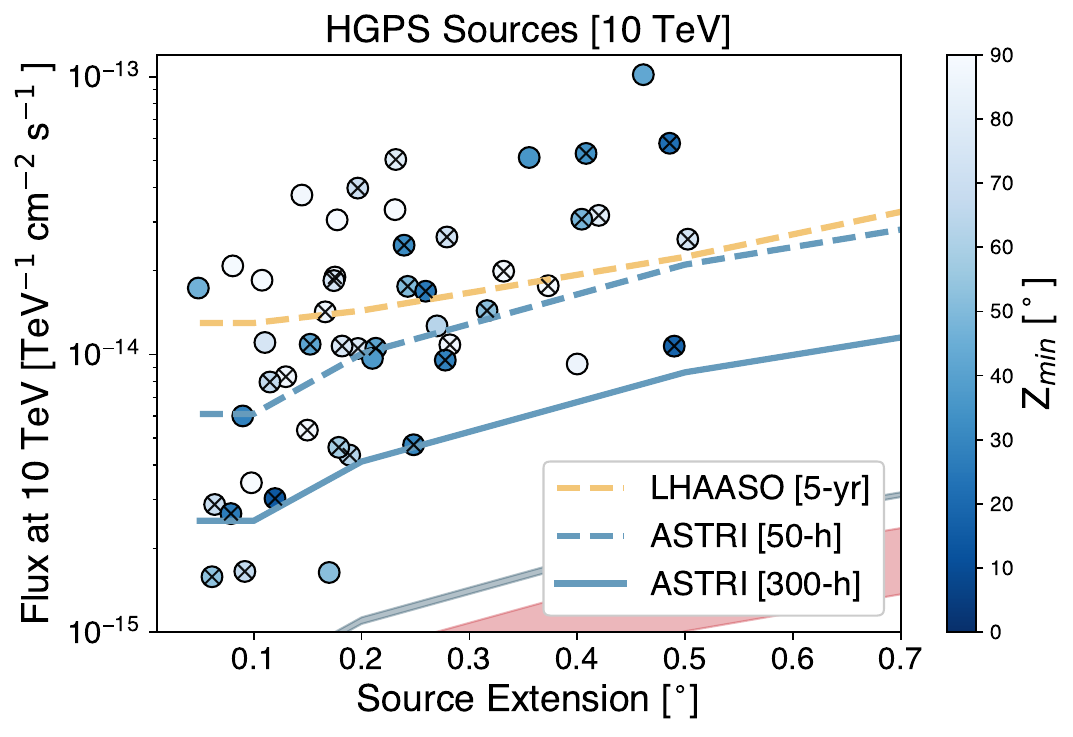}}
    \subfigure[]{\includegraphics[width=0.49\textwidth]{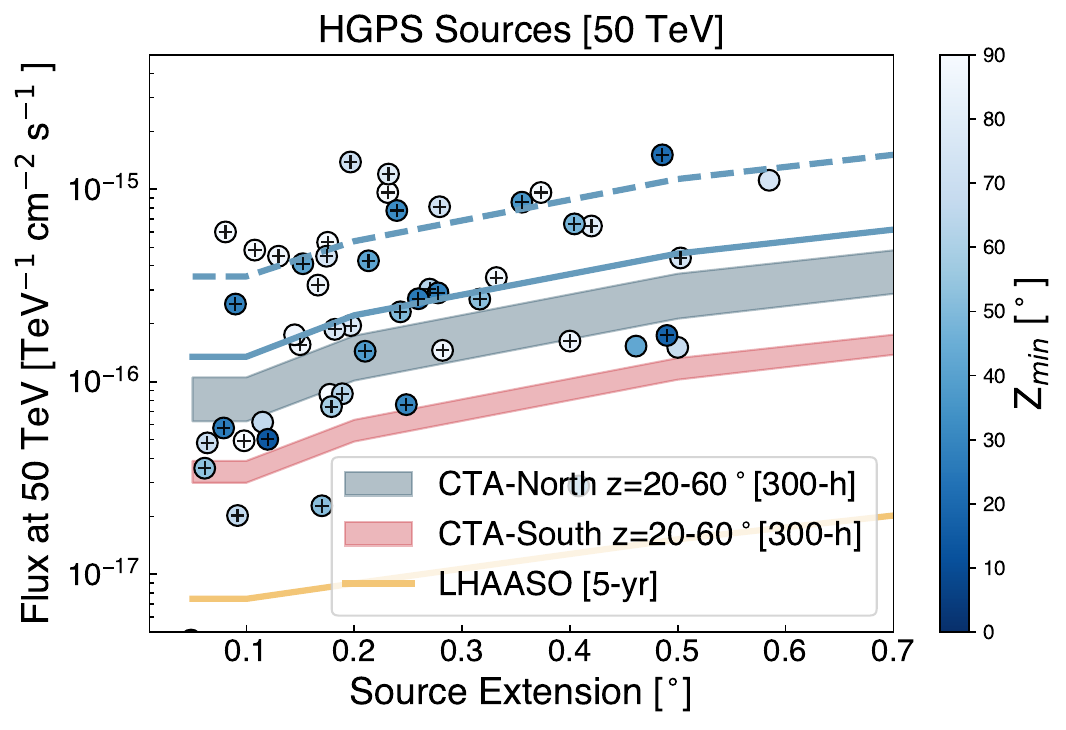}}
    \caption{HGPS \citep{hgps} source fluxes (markers) as a function of their extensions at (a) 10~TeV and (b) 50~TeV, compared to the differential sensitivities (lines) of ASTRI and CTA for different exposures. Sensitivities are evaluated at a zenith of $20^\circ$ for ASTRI, whereas in the case of CTA we span different zenith angles up to $60^\circ$ with a band. In the right panel, source fluxes are extrapolated up to 50 TeV by assuming a power-law or/and exponential cutoff power-law spectrum, according to the indication of the HGPS catalogue. Sources marked with a grey cross in the left panel are unidentified ones, while a plus marker to the right indicates sources without any detected cut-off by H.E.S.S.. The colour code indicates the culmination zenith angle of sources at the ASTRI location, corresponding approximately to that of CTA-N. Note that low-zenith angle observations with CTA-N correspond to high-zenith angle observations with CTA-S. The LHAASO-KM2A differential sensitivity is also reported.} 
    \label{fig:hess}
\end{figure*}

\subsubsection{H.E.S.S. Galactic Plane Survey}
\label{ssec:hgps}
The HGPS \citep{hgps} is a survey performed by H.E.S.S. of the Galactic Plane ($65^\circ<l<250^\circ$, $|b|\leq 3^\circ$) containing 78 sources, detected down to $\sim$ 1\% of the Crab flux. More than 60\% of the sources are unidentified and $\sim$ 80\% are extended. Moreover, only for 15\% of sources (12 objects) energy spectra are consistent with exponential cut-off power laws (a pure power law is fitted otherwise), revealing a maximum cut-off energy in the HGPS at 19.2~TeV. In these respects, it appears clear that observations with next-generation instruments, more sensitive than H.E.S.S. at energies $\gtrsim 10$ TeV, are needed in order to characterise the actual maximum energy achieved by these sources. While we expect CTA to improve the observations of basically all H.E.S.S. detected sources around 10 TeV, it is worth exploring the capability of next-generation IACTs in following up HGPS sources at higher energies.
Despite the sky coverage of the northern observatories and H.E.S.S. is not identical, a few sources are visible from both hemispheres, although culminating at different zenith angles depending on the site. In Fig.~\ref{fig:hess}, we show the HGPS source fluxes at 10 and 50~TeV as a function of the corresponding extension. Note that flux values at 50~TeV are obtained by extrapolating the given spectral slope in the assumption of a pure power-law or an exponential-cutoff power-law with the tabulated parameters for each source. We also show the sensitivity of ASTRI for different exposures at a fixed zenith angle of $20^\circ$ and those of the two CTA observatories with 300~hr observations in a range of zenith angles between $20^\circ$ and $60^\circ$, represented by a band. Moreover, we include a colour scale indicating the zenith angle at culmination for each source at the latitude of ASTRI: in particular the scale takes a white colour for high-zenith angle observations, meaning a less solid comparison with the sensitivity here shown. As a result of the comparison, we find that with a long enough exposure, ASTRI allows almost the entire HGPS sample to be investigated at 10~TeV. Similarly, good prospects are also expected for 50~TeV observations, where more than half of the sources will be in the reach of both ASTRI and CTA. 

Further details about the expected CTA sensitivities with varying zenith are provided in Appendix~\ref{sec:appC}, showing that above few tens of TeV an improvement of about a factor 2 can be achieved in high-zenith angle observations, similarly to what 
was demonstrated in the case of MAGIC \citep{1987JPhG...13..553S,1999JPhG...25.1989K,2020A&A...635A.158M}. We expect analogous considerations to hold for ASTRI, but because of the lack of the official IRFs properly evaluated at different culmination angles, we limit the  discussion concerning ASTRI to low-zenith angles sensitivities. We further note that the zenith angle was evaluated to impact the expected sensitivities of IACTs by an amount that depends on energy differently for CTA-S and CTA-N, as it can be seen in Fig.~ \ref{fig:sens_zenith}. In particular, the variation is quite pronounced for CTA-S at 10~TeV and CTA-N at 50~TeV, while it is more contained for CTA-N at 10~TeV.

\subsection{Identifying hadronic PeVatrons}
\label{ss:pevatrons}
\begin{figure*}
    \centering
    \subfigure[\label{fig:vis_pev_z20}]{\includegraphics[width=0.45\linewidth]{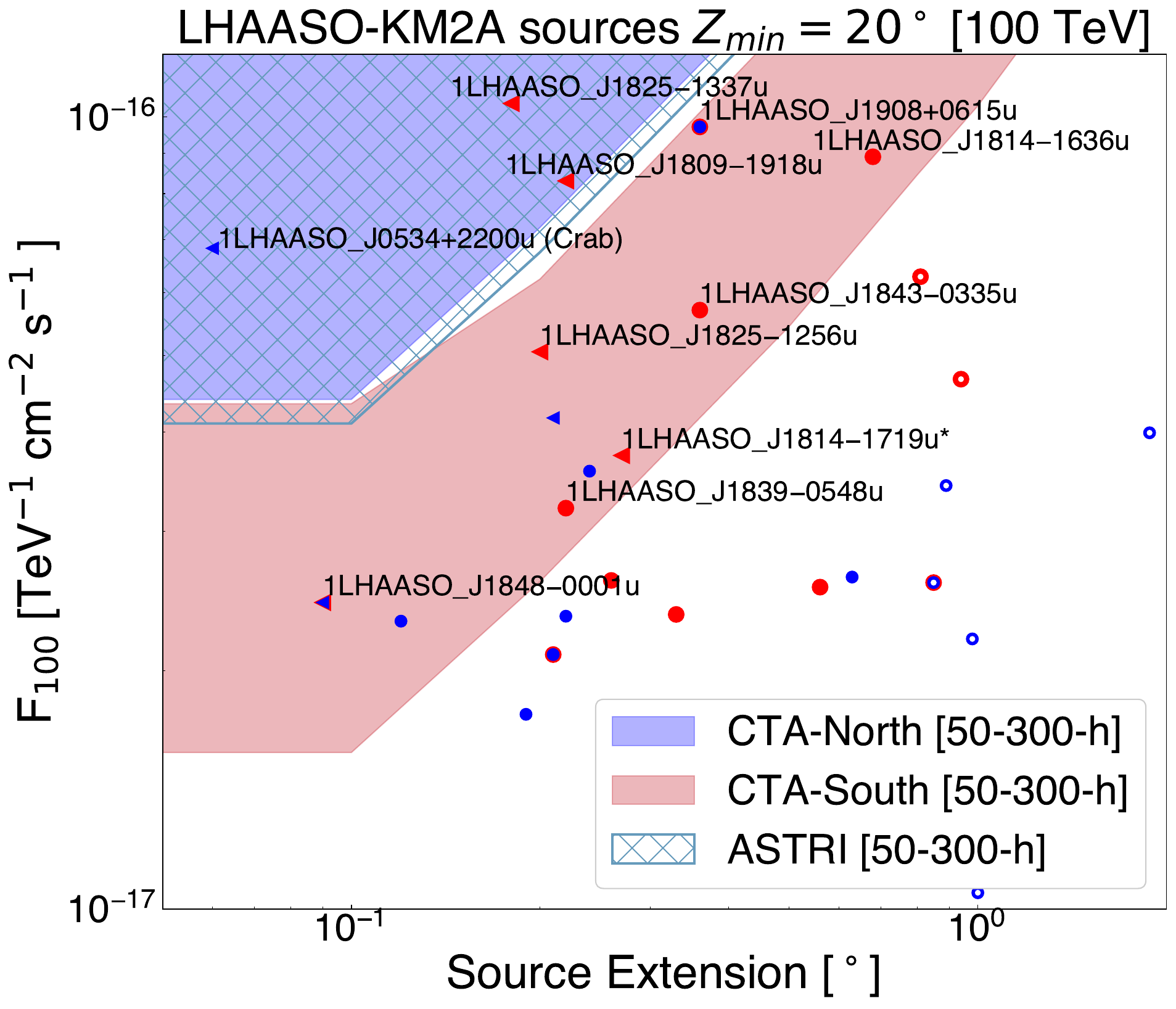}}
    \subfigure[\label{fig:vis_pev_z60}]{\includegraphics[width=0.45\linewidth]{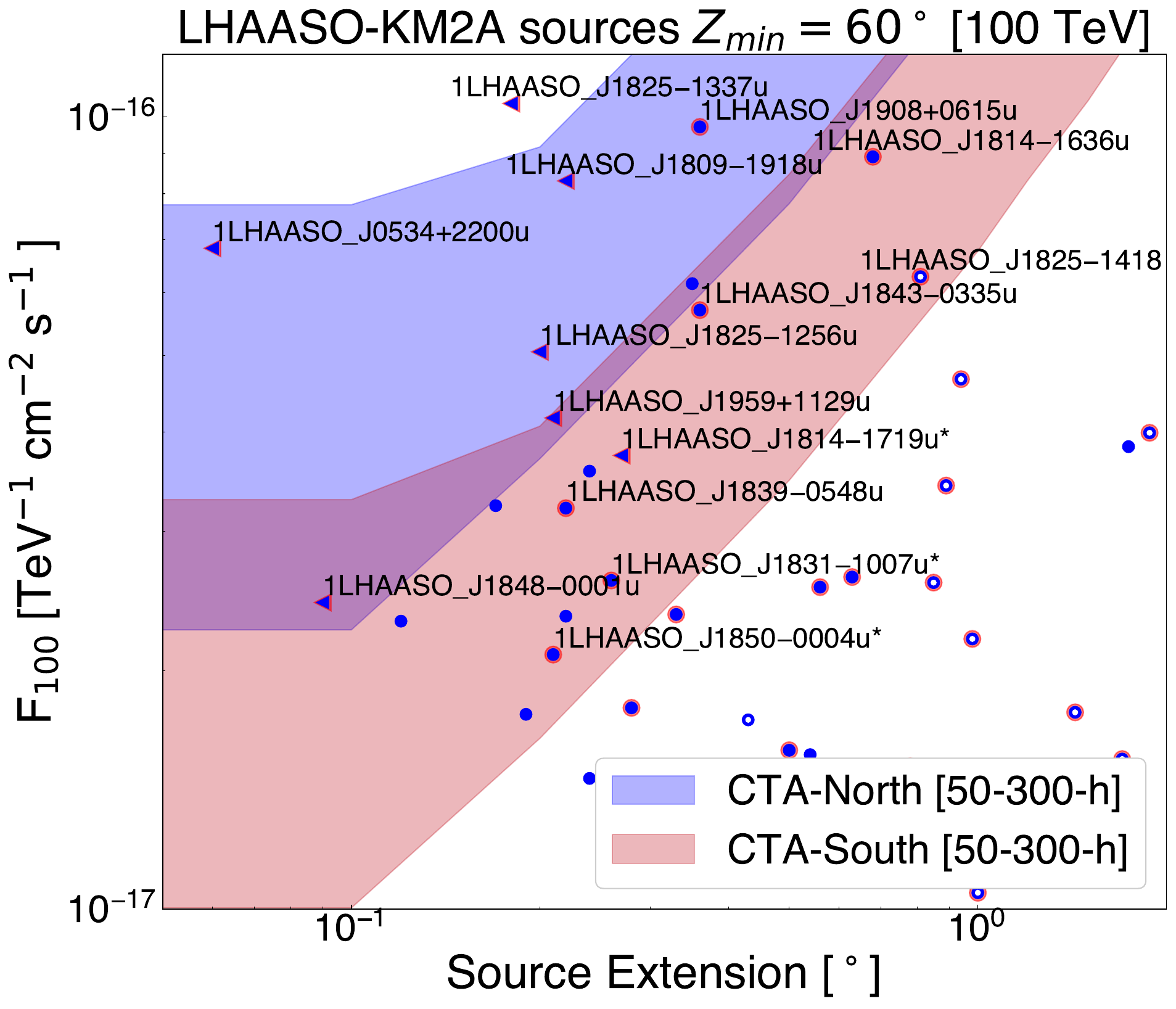}}
    \caption{Expected gamma-ray flux at 100~TeV from the LHAASO-KM2A source sample, as computed by extrapolating the spectral fits provided in \citet{Cao2023Catalog}. UHE sources as identified in LHAASO-KM2A catalogue (${\rm TS}_{100}>20$) are shown in solid markers, while empty markers represent the remaining sample; visible sources by any IACT are labelled with their name identifier in the plot. The marker colour indicates if a sources culminates above the zenith angle $Z_{\rm min}$ (indicated in each panel) in the northern (blue) or in the southern (red) hemisphere. Overlaid are shown the differential sensitivities of ASTRI and CTA, where the bans span different exposure times from 50 to 300 hours. \emph{Left:} The differential sensitivity is calculated with minimum zenith angle of sources equal to 20$^{\circ}$. \emph{Right:} Sensitivities are calculated for minimum zenith angle of 60$^{\circ}$; ASTRI is not included in this panel as its IRFs in such conditions are not yet available for this study.}
    \label{fig:vis_pev}
\end{figure*}

The definition of PeVatron is often considered in the CR and gamma-ray communities with different meanings: in the former, it is intended to address theoretically the first-principle question about the origin of CRs at the knee, while in the other it follows from an observational perspective to describe sources showing signatures of PeV particles, regardless of their contribution to the CR spectrum itself. Hence, when discussing about PeVatrons, it is necessary to clearly state which definition is concerned: in the present case, it will be based on the gamma-ray perspective. Both approaches are, however, model dependent: in fact, in realistic CR theories the maximum energy is often time-dependent, especially in impulsive accelerators as Supernova Remnants (SNRs) \citep{celli2019}, such that detailed models are required to understand whether a given source has ever behaved as a PeVatron; in gamma-ray measurements, a caveat is that the maximum energy derived in fitting procedures strongly depends on the spectral shape of the cut-off \citep{Celli2020Pevatron}, such that solid conclusions can only be derived as a result of data descriptions with regards to a given model, not simply by minimising the residuals with respect to a standard functional form.  \\
With respect to the standard paradigm of Galactic CRs suggesting SNRs as main CR sources, we highlight that so far LHAASO observations have revealed a few UHE emitters \citep{Cao2023Catalog} in spatial coincidence with known SNRs, all sources being characterised by featureless and steep gamma-ray energy spectra. While the collection of increased statistics and multi-wavelength analysis are necessary to clearly assess the overall spectral trend of their emission and to probe the presence of spectral variations, it is unlikely for these objects to behave as main CR contributors. 
In these regards, the improved energy resolution will be a key feature of future IACTs, possibly providing the resolving power to distinctly identify spectral features at the highest energies. The list of LHAASO detected sources include many other objects showing significant emission $>100$~TeV. Among these, the only clear identification so far concerns the Crab Nebula, mostly dominated by leptonic emission, although a hint for a spectral hardening at the highest energies probed by LHAASO suggests an additional component possibly be due to hadrons \citep{nie2022}. The confirmation of this scenario awaits improving mapping of the emitting region at these extreme energies.  \\
Given the primary importance to investigate which of the LHAASO detected sources can be observed at 100 TeV by the next-generation IACTs, both in the context of exploring spectral features and associate unidentified ones, we now focus on the KM2A detected sources from the first LHAASO catalogue. As no information is available in the catalogue concerning the maximum photon energy measured for each of them, we extrapolate the reported fluxes up to 100 TeV. In Fig.~\ref{fig:vis_pev} we compare the resulting flux at 100 TeV, $F_{100}$, to the differential sensitivity of ASTRI and CTA at the very same energy: in particular, the left panel refers to observations at $20^\circ$ zenith angles, while the right panel concerns high-zenith angle ($60^\circ$) observations for CTA only. As it emerges from our estimation, CTA-South has the potential for investigating several objects at 100~TeV; among these, the source powering the region of 1LHAASO 1825-1337u and 1LHAASO J1809-1928u, possibly linked to pulsar emission \citep{Cao2023Catalog}, will be potentially visible already with a short 50-hour exposure. A deeper exposure coupled to high-zenith angle observations will enable all observatories to access a larger number of sources, despite their extensions. All detectable sources are labelled with their name in Fig.~\ref{fig:vis_pev}. \\

\subsection{Molecular clouds} 
\label{ss:mc}
Molecular clouds (MCs) illuminated by CRs constitute a class of Galactic gamma-ray emitters, typically detected in the GeV energy range \citep{Aharonian2020Clouds, Peron2021Clouds,Vardan2020}. These are compact ($\sim 10-100$~pc) regions of enhanced gas density ($n\sim 10-100$~cm$^{-3}$) recognizable in the surveys of CO (J=2$\rightarrow$1) line \citep{Dame2001}. When high-energy CRs impact the clouds, they produce (among other secondaries) gamma radiation proportional to both the density of the target gas and that of CRs, making MCs the most straightforward tracers of CRs far from the Earth. \\
In the following, we evaluate the cloud visibility of the MCs listed in the catalogue of \citet{Miville2017} for the next-generation instruments in the multi-TeV energy range, by comparing the expected gamma-ray flux emerging from hadronic collisions of the local CR flux with their sensitivity, accounting for the cloud extension and the column density reported in the catalogue. Following \citet{PeronAharonian2022}, we compute the resulting photon flux by assuming the locally measured CR flux $J_{\odot}(E_{\rm p})$ \citep{Orlando2018CRs,Lipari2020}.
The assumption that the same CR spectrum across the Galaxy illuminates the clouds implies that the resulting hadronic gamma-ray flux only depends on their mass and distance. Indeed $F_{\rm cloud} \propto n_{\rm col}~d\Omega = M/d^{2}$, in specific units. Following \citep{Aharonian2020Clouds}, we define $A= M_5/d^{2}_{\rm kpc}$ as the ratio between the cloud mass (normalised to 10$^{5}$ solar masses) and the squared distance (in kpc units), and factorise the expected gamma-ray flux as $F_{\gamma}(E) = A \phi_{\gamma}(E)$, being $\phi_\gamma(E)$ the emissivity per single hydrogen atom. As a consequence, the visibility condition translates into a mere constraint regarding $A$: defining the instrument sensitivity for given source extension as $S_{\rm min}(E,R_{\rm src})$, then 

\begin{equation}
F_{\gamma} (E) > S_{\rm min}(E,R_{\rm src}) \rightarrow A> S_{\rm min}(E,R_{\rm src})/\phi_{\gamma}(E).
\end{equation}

\begin{figure*}
    \centering
    \subfigure[]{\includegraphics[width=0.49\textwidth]{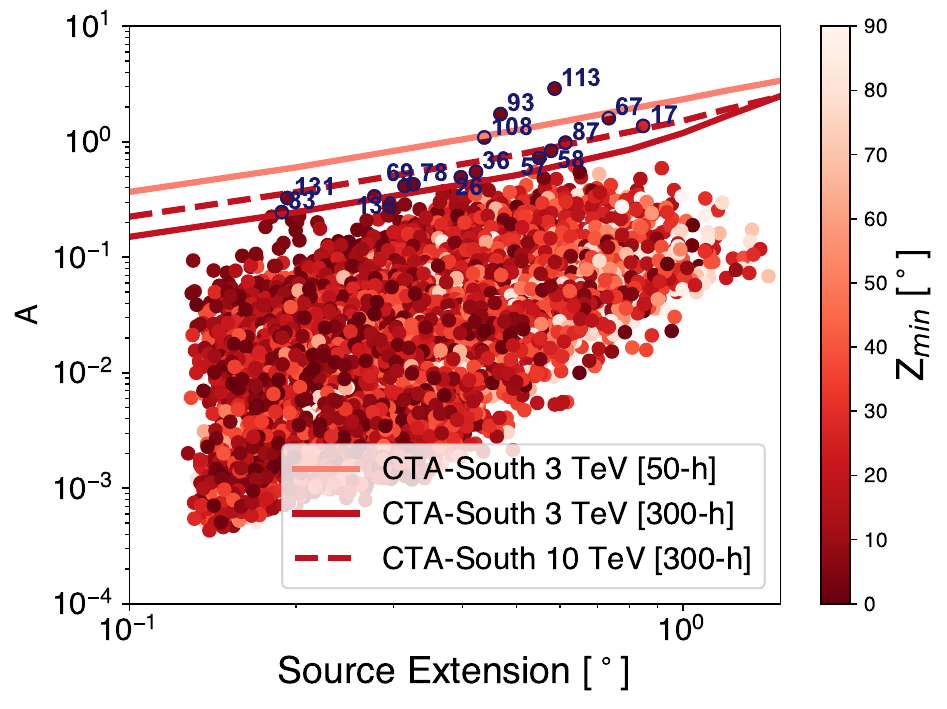}}
    \subfigure[]{\includegraphics[width=0.49\textwidth]{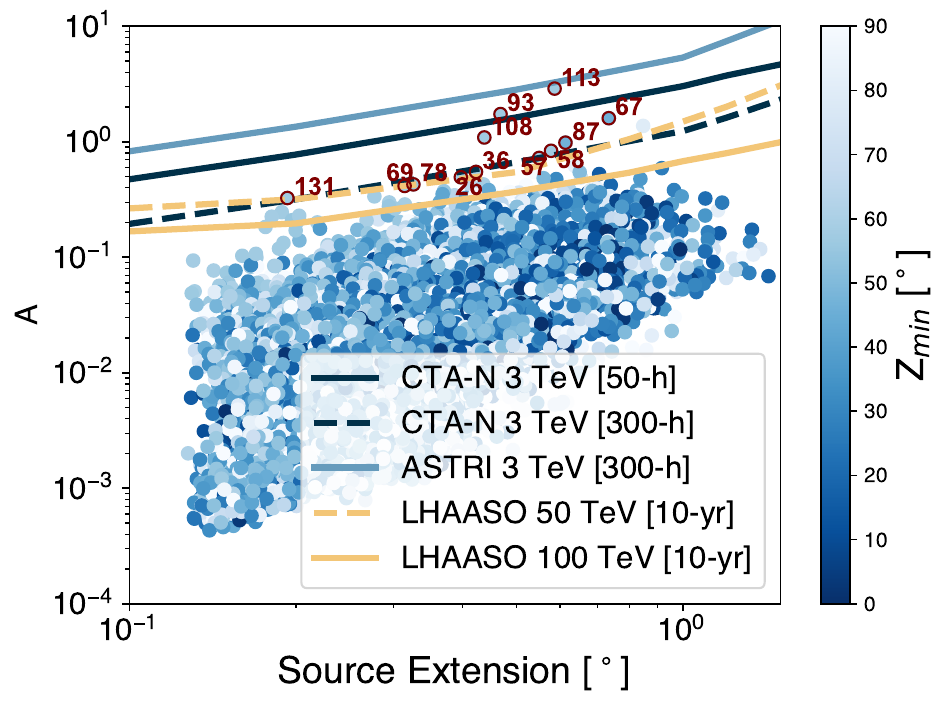}}
    \caption{Detectability of MCs from the \citet{Miville2017} catalogue (circles), in terms of their computed $A$ parameter and extension. The sensitivity of various instruments at different energies are normalised to the expected cloud emissivity at the same energy (see text). The left panel refers to CTA-S, while the right panel to northern observatories. The same cloud sample is reported in the two panels, while the colour scale refers to the minimum zenith angle of the clouds at each location. Clouds with expected gamma-ray fluxes above the sensitivity requirements and culminating at a zenith angle smaller than 60$^\circ$ are further labelled in the figure with their identifier, as defined in the \citet{Miville2017} catalogue.}
    \label{fig:MC_vis}
\end{figure*}

In Fig.~\ref{fig:MC_vis} we compare the $A$ parameter characterising the catalogued MCs to the ratio between sensitivity and emissivity at various energies and for different exposures. Also in this case, the zenith angle at which the cloud is observed might influence the detection, especially at low energies. We consider here only the case of low-zenith angle observations ($< 20^{\circ}$), and we indicate through the colour code of each cloud the zenith angle of their culmination at the southern and northern sites of CTA, the latter being located approximately at the same latitude as ASTRI and LHAASO. Potentially detectable clouds are highlighted with a label reporting their unique identifier as provided in the \citet{Miville2017} catalogue. To this extent, we selected clouds with an expected gamma-ray flux overcoming the detection threshold for extended sources with at least one of the considered IACTs and culminating at zenith angles lower than 60$^\circ$ at the corresponding site. As it can be seen from the plot, the clouds with highest $A$ culminate at lower zenith angles in the southern hemisphere, resulting in a possible detection of a few targets already with a 50-hour exposure by CTA-S. In northern sites, on the other hand, the same clouds culminate at high-zenith angles, where observations are characterised by an increased energy threshold (see Appendix~\ref{sec:appC}). This is however limiting the prospects for the observation of the MCs that are illuminated by local CRs only, because a steep gamma-ray emissivity is expected in this case.
Concerning LHAASO, a few clouds are found to be possibly detectable only with a 10-year long exposure. As a consequence, our result clearly suggests that passive MCs illuminated by the CR sea cannot be responsible for the unassociated sources in the first LHAASO catalogue \citep{Cao2023Catalog}, which is based on a 933-day long exposure. However, future detection of MCs could enable the Galactic CR spectrum to be probed up to $> 100$~TeV, which is a particularly interesting energy range because of the discrepant results currently shown by ground-based measurements \citep{icetop}.

For a few of these clouds, namely cloud 57 and 78 the \textit{Fermi}-LAT analysis \citep{Peron2021Clouds} unveiled a spectrum similar to the locally measured one. Nevertheless, one should consider that a large level of enhancement in the CR flux is expected in the proximity of accelerators, due to the escaping particles \citep{aharonian1996,gabici2007}. E.g., the scenarios where Galactic clouds could be illuminated by SNR and YMSC escaped particles were studied in great details by \citet{Mitchell2021SNRClouds} and \citet{Celli2023SCClouds}, respectively. Furthermore, many among the clouds observed in gamma rays showed an enhancement as large as $\sim$5 compared to \emph{local} (i.e. at Earth) CR measured flux, even if not directly associated with an accelerator: the increased emissions from clouds were found mostly in the inner part of the Galaxy \citep{Aharonian2020Clouds}, but also in the local ($\lesssim 2$ kpc) medium \citep{Vardan2020,Peron2023icrc}. Similar enhancements would dramatically increase the detection prospects, allowing us to probe cloud spectra even at high energies. One compelling case is the case of cloud 17, which coincides with the cloud 877 of the \citet{Rice2016clouds}. Its enhanced emission reported by \textit{Fermi}-LAT \citep{Aharonian2020Clouds} stimulated follow-up observations with H.E.S.S.  \citep{Sinha:2021icrc}, that reported a clear detection and proposed a dedicated analysis method for such low-surface brightness sources. 
The extension of these type of observations to other objects, although challenging, would be of extreme relevance to probe the CR density elsewhere in the Galaxy especially at high energies, where a clear picture is not yet there. 

\subsection{PWNe and pulsar halos}
\label{ss:pwn}
\begin{figure}
    \centering
   \includegraphics[width=1 \linewidth]{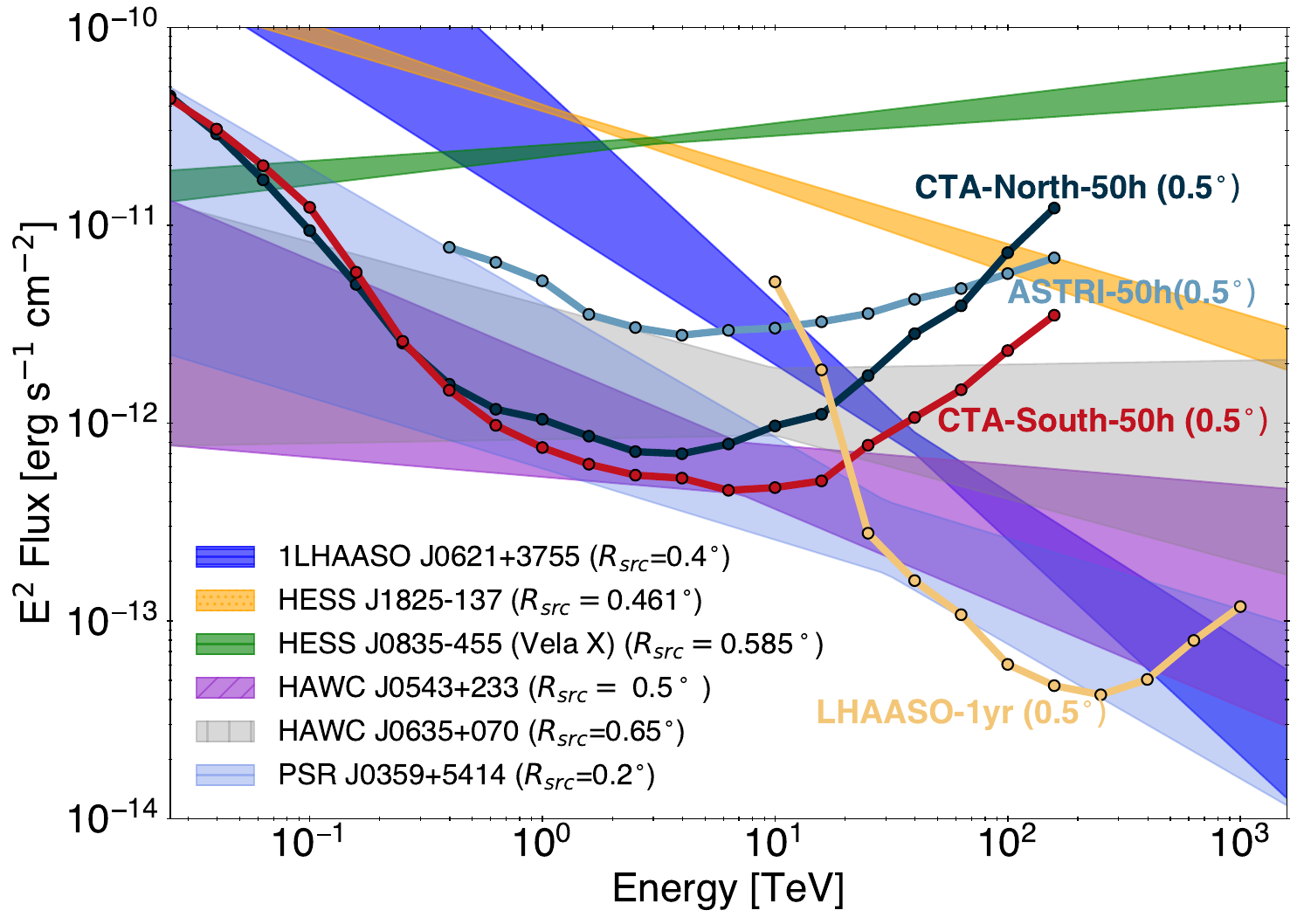}
    \caption{Differential flux of the candidate TeV halos discussed in the text, compared with the differential sensitivity for extended ($R_{\rm src} = 0.5^\circ$) sources of the next-generation IACTs. The LHAASO-KM2A differential sensitivity is also reported, showing its capability of unveiling all of the hereby investigated sources provided these are in its field of view: specifically this is the case for the HAWC candidates and PSR J0359+5414. The measured size of each source is also reported in the legend, as extracted from TeVCat. The shaded bands are computed from the $1\sigma$ spectral uncertainties (slope and normalisation) also provided there.}
    \label{fig:vis_halos}
\end{figure}

A new source class, named pulsar/TeV halos, has  emerged recently in the VHE surveys of the large FoV instrument HAWC. These sources are found in correspondence of pulsars, despite being much more extended than the associated surrounding PWN, therefore interpreted as a halo of particles escaping from the nebula.  Interestingly, LHAASO observations have revealed that 40\% of the first  catalogue sources are coincident with pulsars, possibly powerful enough to support UHE emission \citep{emma}.\\
The first detection of such objects was obtained in the direction of the Geminga and Monogem pulsars \citep{hawcTH}: around these pulsars, a $\gtrsim 2^\circ$ halo was measured by HAWC, corresponding to a distance of about 50 pc from the central compact object. 
Current IACT observations are limited by the extension of such halos, covering (and often going beyond) their entire FoV as in the case of Geminga (\cite{Hess2023Geminga}), representing a noticeable analysis challenge for the background estimation. Furthermore, significant flux differences have emerged in current measurements from IACTs and EASs of several sources because of the different integration region, making the case for observations with large FoV IACTs, such as ASTRI, more compelling than ever. \\
Interestingly, in order to explain the halo emission, a strong suppression of the diffusion coefficient or a transition to a different propagation region \citep{Recchia2021halo}, is required, with implications regarding the amount of CR positron produced by these sources and in general on the propagation of Galactic CRs \citep{ruben}. However, Monogem and Geminga are found in the same region, therefore a local deformation of the magnetic field, which would impact the diffusion of particles regardless of the presence of the pulsars, cannot be excluded a priori. Despite this hypothesis being disfavoured theoretically \citep{Delatorre2022halos}, further observations are necessary to conclude about the origin of the observed halos and to which extent they can influence the diffusion properties of the interstellar medium. \\
At the moment, it is not clear whether these sources are the only pulsar halos in the sky or if any other exists, although some evidence for similar phenomena emerged in the most up-to-date catalogues of TeV sources. One candidate pulsar halo is the source HESS~J1825-137, because its extension appears significantly larger than ordinary PWNe, suggesting that escape must be taking place from the nebula. Another candidate is LHAASO~J0621+3755, one of the first sources detected in UHE gamma rays, its morphology being compatible with the scenario of escaped particles as claimed in \citet{Aharonian2021PWNe}. The observations with LHAASO-KM2A point towards a suppression of the diffusion coefficient to a similar level as measured in Geminga and Monogem. Again, high resolution spectral and morphological observations would be crucial for confirming the nature of this accelerator. Other candidate TeV halos have been announced by the HAWC Collaboration, namely HAWC~J0543+233 and HAWC~J0635+070, both included in the second HAWC catalogue \citep{Abeysekara2017Hawccat}: also in these cases, further morphological and spectral characterisation are needed to be confirmed as halos, as current measurements could not resolve the radial profile of the emission neither the energy cut-off region with sufficient accuracy. We note that energetic emission is measured by LHAASO from a direction consistent with both these sources, namely LHAASO J0543+2311u and LHAASO J0635+0619, both observed by KM2A the former even at UHE. All of the aforementioned sources are expected to be powered by middle-age pulsars ($\gtrsim 100$ kyr). An exception is found in the pulsar of Vela, namely Vela X: Vela shows a dim extended emission detected both in radio and in the high-energy range, which is often interpreted as 'relic PWN', left behind after the interaction of the expanding PWN and the SNR reverse shock. The extension of the emission is however compatible also with a TeV halo scenario, making Vela X a possible intermediate case between the relic and the halo phase \citep{giacinti2020}. Similarly, a TeV halo interpretation has been suggested to explain the extended emission detected by  HAWC around the young pulsar PSR J0359+5414 \citep{kefang2023}. Such a pulsar, also detected by \textit{Fermi}, does not show a radio counterpart, analogously to the case of LHAASO J0621+3755. If this association is confirmed, it would imply that a larger variety of pulsars, compared to the present understanding, is able to produce halos. The spatial association with LHAASO J0359+5406 remains to be confirmed, given that also PSR B0355+54 could in principle power its emission. \\
For these candidates, we provide in Fig.~\ref{fig:vis_halos} a comparison between their differential fluxes and the calculated sensitivity of future IACTs and LHAASO for a 0.5$^\circ$ extension, compatibly with their reported sizes in TeVCat\footnote{\url{http://tevcat.uchicago.edu}} and reported in the figure legend. The selected sources will be possibly visible with the benchmark exposures by the planned instruments, therefore allowing a comprehensive characterisation of the emission around these pulsars, that so far is still missing. Investigating with high resolution the spatial profile of the flux of runaway particles might yield insight into radially dependent diffusion properties. Such a goal is expected to be achieved in the future: in fact, the limited performance worsening when moving away from the axis centre (Figs.~\ref{fig:astri_offset_IRFs}, \ref{fig:ctan_offset_IRFs}, \ref{fig:ctas_offset_IRFs}, and \ref{fig:offset_sens}) can be favourably exploited in such an application. We therefore expect high quality results in this field, although a more detailed evaluation needs to be performed to quantify the detectability of very large emitting regions as Geminga, e.g. accounting for the exact source geometry and spatial dependency of the spectrum. 

\section{Discussion}
\label{sec:discussion}
\begin{figure*}
    \centering
    \includegraphics[width=1\textwidth]{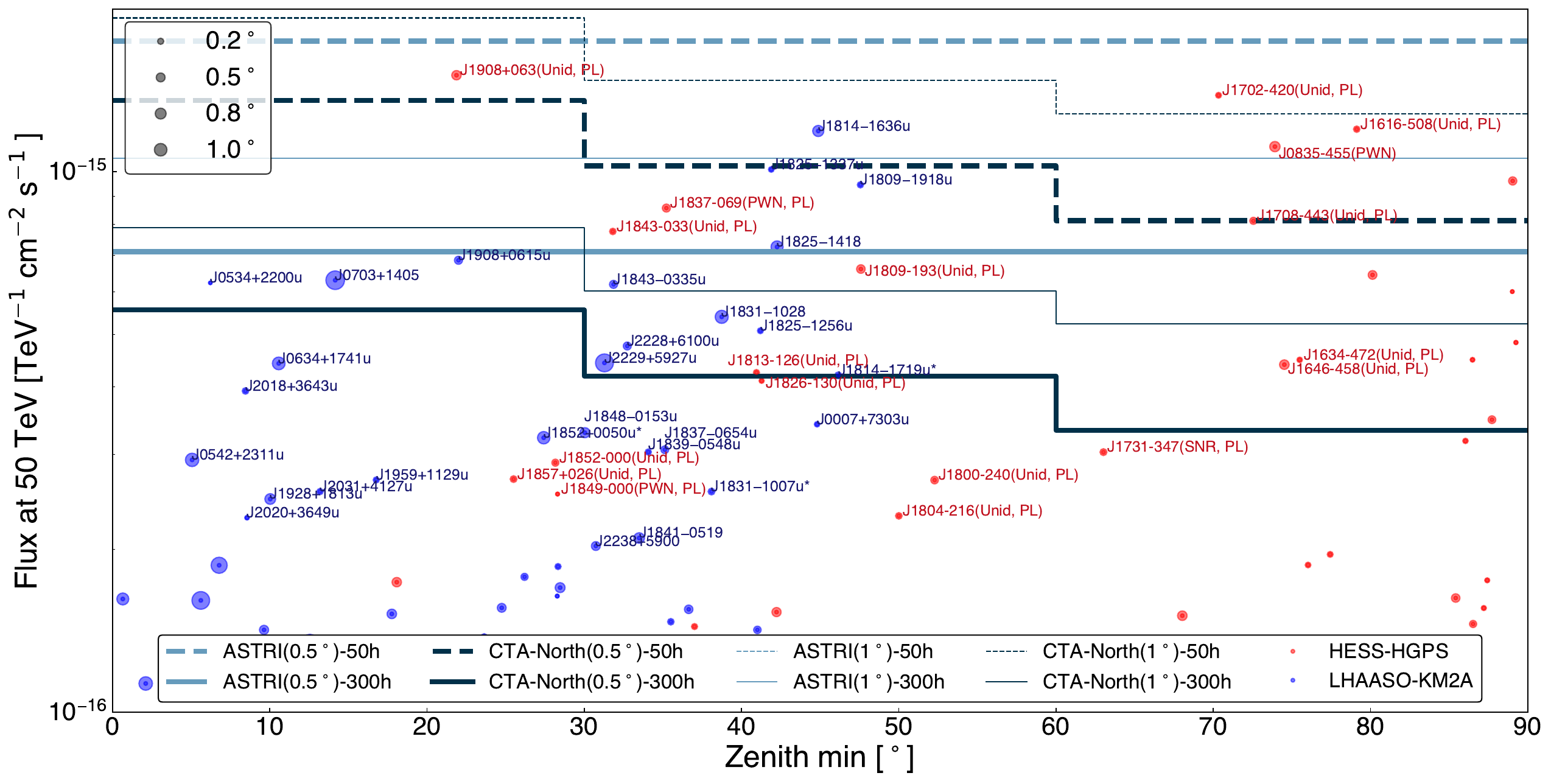}
     \includegraphics[width=1\textwidth]{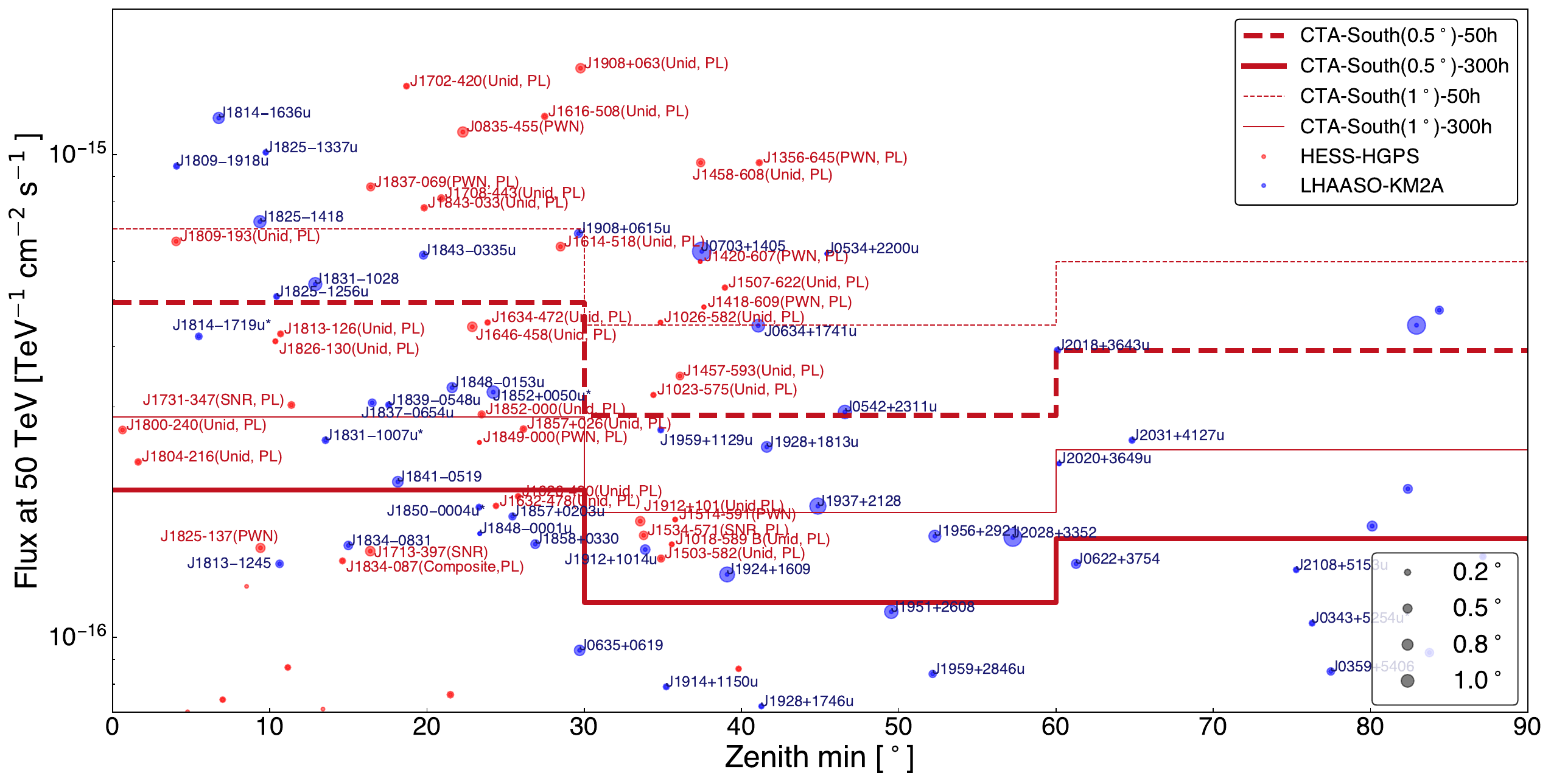}
    \caption{Differential gamma-ray flux at 50~TeV, either measured from LHAASO-KM2A source sample (blue dots) or extrapolated from the H.G.P.S. catalogue (red dots), as discussed in the text, as a function of the minimum zenith of observation from the northern sites (left panel) and the southern sites (right panel). Differential sensitivity lines of IACTs are also shown for comparison at different exposures and for extended source observations, as indicated in the legend.}
    \label{fig:catalogs_unid}
\end{figure*}

In the light of defining the best candidates for future IACT observations, we provide in Fig.~\ref{fig:catalogs_unid} gamma-ray source fluxes at 50~TeV from both the LHAASO-KM2A and the H.G.P.S. catalogues, as compared to sensitivities of IACTs. 
The choice of 50 TeV as a reference energy follows from the measurements by LHAASO-KM2A sources. On the other hand, for H.E.S.S. sources we obtain the expected 50~TeV fluxes by extrapolating the measured spectra. Given the impact of the zenith angle on IACT sensitivity, we show source gamma-ray fluxes as a function of the minimum observation angle in the northern and southern sites, respectively in the upper and lower panels of the figure. We expect that the major observational efforts will be most-likely devoted, in the next decade, to unveil the nature of unidentified sources and measure possible cut-off features to find out any source that does not show a cut-off earlier than PeV. To facilitate the identification of these objects, we explicitly indicate in the figure, next to source names, their class (unidentified, SNR, PWN, binaries, etc.) as well as their best-fit spectral shape (PL, where a cut-off is not detected), whenever this information is available. 
We can hence define a list of the most promising targets for each hemisphere, selecting the top five brightest unidentified sources characterised by featureless spectra, observable at zenith angles below 60$^\circ$ in the corresponding sites. 
For the northern sites, we highlight: \\
\begin{itemize}
\item HESS J1908+063/1LHAASO J1908+0615u: these two spatially overlapping sources have been revealed by different surveys, but it remains to be clarified if they are the same or not. We note that the flux measured by LHAASO-KM2A at 50 TeV is much lower than the extrapolated flux from H.E.S.S., therefore if those are indeed the same source, a spectral break  would be necessary to connect the two spectra. Multiple counterparts are possible PWN and SNR but also the Juchert 3 star cluster \citep{hess1908,celli2023}
\item 1LHAASO J1814-1636u: this UHE source is only detected by KM2A with angular extension of $\sim 0.68^\circ$. Two SNRs are located nearby, namely G14.1-00.1 ($\sim 0.43^\circ$ away) and SNR G014.3+00.1 ($\sim 0.31^\circ$ away). However, both SNR candidates have $< 0.2^\circ$ radio shell size, hence inconsistent with that of gamma-ray emission. An unidentified GeV source, 4FGL J1816.2-1654c, is also found $0.4^\circ$ away from the position of the TeV emission, but so far no clear counterpart emerged.
\item HESS J1825-137, 1LHAASO J1825-1337u, and 1LHAASO J1825-1418 (good targets also for the southern observatories): the H.E.S.S. source is spatially overlapping with two LHAASO sources, though the association is unclear. Notably in this case, the source observed by H.E.S.S. has a flux significantly smaller than the LHAASO sources, such that it again appears necessary to determine whether the emission can be ascribed to the same source or if multiple emitters are present. In both cases, an emerging component at 50 TeV is particularly interesting. The H.E.S.S. sources, identified as PWN by the collaboration, is one of the candidate TeV halos that we discussed in Sec.~\ref{sec:science}.
\item HESS J1809-193/1LHAASO J1809-1918u: analogously to the previous case, also here, two apparently coincident sources have an inconsistent spectrum, where the LHAASO measured flux is much larger than the one extrapolated from H.E.S.S; in this case, though, the counterpart remains obscure. In the past, this source was thought to be a PWN, but other possible counterparts such as SNRs and molecular later emerged, making the recognition of the dominating emission mechanism not straightforward. \citep{hess1809}. 
\item HESS J1843-033/1LHAASO J1843-0335u: little is known about this source, that seems to be characterised by two components; in particular, the H.E.S.S. source coincides with UBC~353, a weak and quite old star cluster from the \textit{Gaia} catalogue \citep{celli2023}.
\end{itemize}
For the southern sites, in turn, we suggest in-depth observations to be carried for:
\begin{itemize}
\item HESS J1702-420: this unidentified source is a candidate PeVatron, due to its hard energy spectrum up to the highest energies probed \citep{hess1702}, such that it is among the most interesting sources to investigate.
\item HESS J1616-508: this source is also unidentified in the H.E.S.S. catalogue and, due to the conspicuous nature of the diffuse gas in there, it is possibly composed of a yet-undetected central accelerator producing a particle flux interacting with the gas \citep{hess1616}.
\item HESS J1458-608: with regards to this unidentified source, its VHE gamma-ray emission is thought to be associated with the pulsar PSR J1459-6053 such that it seems likely to be an undetected PWN, on the basis of a spatial coincidence with an energetic pulsar and the absence of other plausible multi-wavelength counterparts \citep{hess1458}.
\item HESS J1708-443:  this is a faint extended source that is possibly associated to a PWN, although a SNR is also a possible counterpart \citep{hess1708}.
\item HESS J1614-518: described as an 'SNR candidate' with a shell-type morphology \citep{hgps}, it remains to date without any multi-wavelength counterpart, despite  extensive campaigns \citep{hess1616}.
\end{itemize}

\section{Conclusions}
\label{sec:conclusion}
The highest energy window ever probed in gamma-ray astronomy has recently been opened by LHAASO observations, reporting extended and often complex emitting regions located in the plane of the Galaxy. The limited angular resolution of EAS arrays has so far impeded the clear identification of the nature of these objects, which represents one of the next major goals of the VHE community. Improved morphological measurements by next-generation IACTs are expected to reduce source confusion from nearby objects. At the same time, lowering the angular resolution will allow us to perform significantly better correlation studies between gamma-ray and multi-wavelength data, which are very relevant to unveil the origin of the gamma radiation. Moreover, accurate spectral analyses might provide further insights into the physical mechanisms at the origin of the radiation. Follow-up analyses with IACTs on LHAASO sources will help to improve the spectral modelling and therefore the source understanding overall.

In order to get a realistic evaluation of IACT detection prospects and required exposure times with regards to known extended gamma-ray sources, we have performed a comparative analysis of ASTRI, CTA, and LHAASO. Starting from their expected performances, we defined a method to obtain differential sensitivities towards the observation of extended sources. We found minimum detectable fluxes at the level of $10^{-12}$~erg~cm$^{-2}$~s$^{-1}$ for a 50~hr exposure by CTA-N (at 3~TeV), $5 \times 10^{-13}$~erg~cm$^{-2}$~s$^{-1}$ for a 50~hr exposure by CTA-S (at 8~TeV), $3 \times 10^{-12}$~erg~cm$^{-2}$~s$^{-1}$ for a 50~hr exposure by ASTRI (at 3.5~TeV), and $6 \times 10^{-14}$~erg~cm$^{-2}$~s$^{-1}$ for a 1~yr exposure by LHAASO (at 100~TeV); all  were evaluated with disk-like sources with an angular radius equal to $0.5^\circ$ and at  $20^\circ$ zenith angle observations. We further explored the expected IACT detection improvements for high zenith angle observations at the highest energies.

Since our analysis relies on publicly available IRFs provided by each collaboration, our method to assess the sensitivity for extended sources includes neither any optimisation of the tools for the reconstruction of the properties of primary particles, nor dedicated background rejection methods. In addition, a precise estimation of the instrument capabilities for the detection of extended sources would require a complete spectro-morphological analysis that goes beyond the simplified photon-counting approached adopted here.

Moreover, we considered a single shape for the source, a uniform disk, but our methods can in principle be extended to any source morphology, including the case of asymmetric shapes for which we expect some differences as a consequence of non-uniform responses of the instruments across the FoV. For IACTs, the study of sub-structures on arc-minute scales would permit the morphological details of many extended sources to be resolved beyond the disk-like structure considered here. Future studies will deal with the contamination of nearby sources as well as the determination of the contribution of the Galactic diffuse emission as an additional background source.

Our investigation enabled us to identify the  most promising  extended targets, considering both the first LHAASO catalogue and HGPS, that are to be observed with the next generation of IACTs. We remark that the detection assessment derived here should be regarded as rather conservative because it resulted from considering region of interests of the same extent as that announced by LHAASO; on the other hand, if the actual source size were smaller than claimed, a detection from IACTs may require a shorter exposure than what has been estimated here. In particular, IACTs are expected to be sensitive to emission peaks, which might be spread out in EAS observations.

We discussed specific key science cases to resolve: i) candidate PeVatrons, potential CR sources; ii) passive molecular clouds, potential gamma-ray unidentified sources; and iii) TeV halos, a novel class recently observed in gamma-ray astronomy. Additional studies with future IACTs might shed light on each of these open scientific questions.

\begin{acknowledgements}
This research has made use of the CTA instrument response functions provided by the CTA Observatory and Consortium, 
see \url{https://www.cta-observatory.org/science/cta-performance/} (version prod5 v0.1; \url{https://doi.org/10.5281/zenodo.5499840}) for more details, and of the ASTRI Mini-Array Instrument Response Functions (IRFs) provided by the ASTRI Project \citep{astri2022IRFs}. The authors thank F. Aharonian, S. Gabici, E. Amato, C. Zhen, S. Zhang, C. Li, and W. Zha for fruitful discussions. SC acknowledges fundings from Sapienza University of Rome with grant agreement RG12117A87956C66. The research of GP was partly funded by  Agence Nationale de la Recherche (grant ANR-21-CE31-0028). GP further acknowledges financial support from the INAF initiative "IAF Astronomy Fellowships in Italy". This paper went through the internal ASTRI and CTA review processes, for which we acknowledge comments from A.~Giuliani, S.~Lombardi, and M.~Strzys that have improved the manuscripts.
\end{acknowledgements}

%
\bibliographystyle{aa} 
\bibliography{biblio} 
%

\begin{appendix}

\section{Integral sensitivities}
\label{sec:appA}
\begin{figure*}
    \centering
    \subfigure[]{\includegraphics[width=0.49\textwidth]{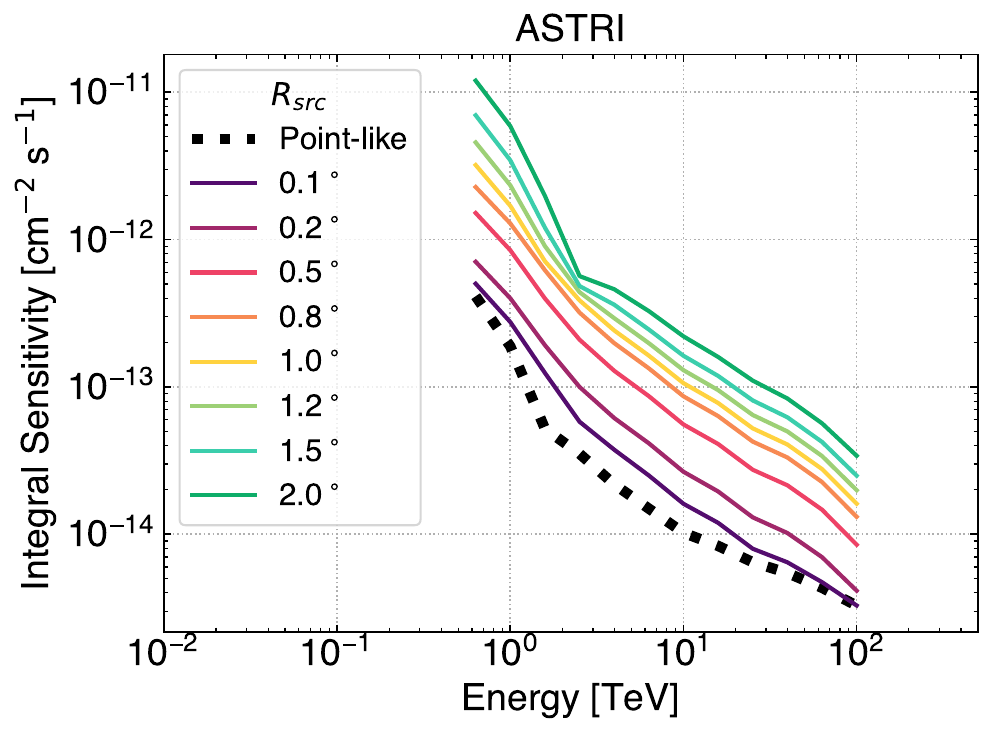}}
    \subfigure[]{\includegraphics[width=0.49\textwidth]{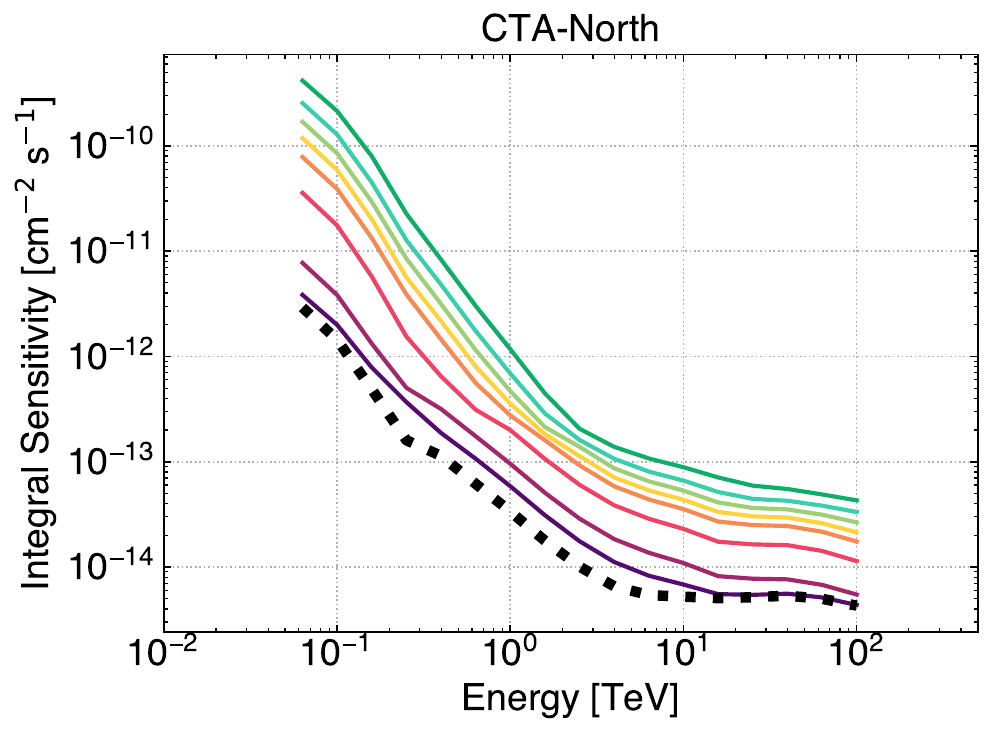}}
    \subfigure[]{\includegraphics[width=0.49\textwidth]{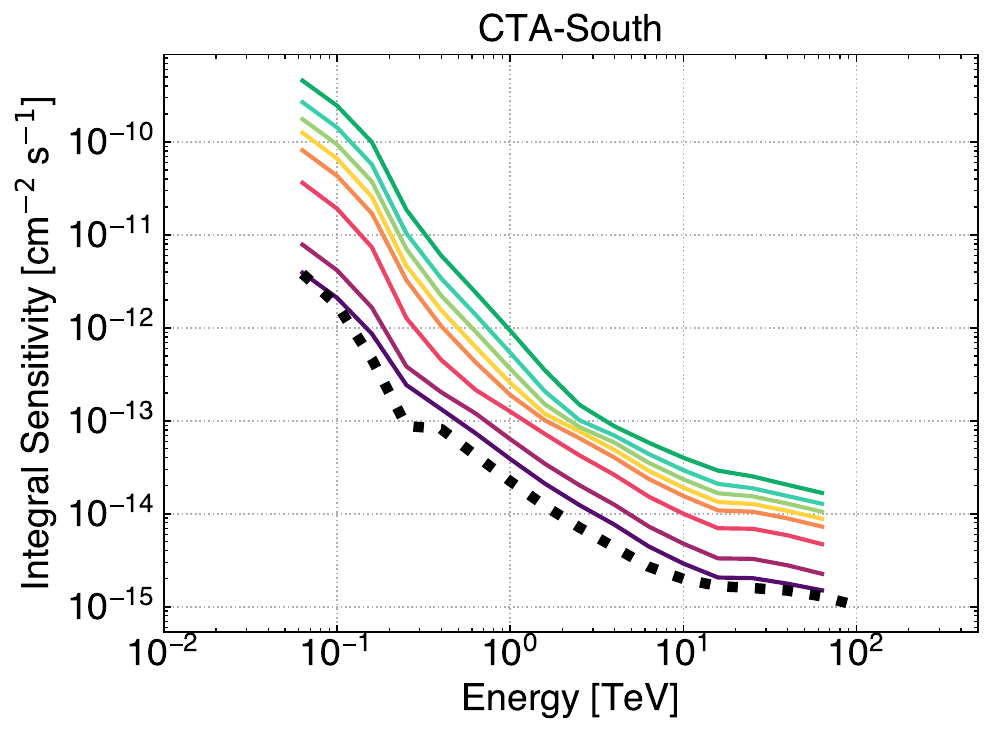}}
    \subfigure[\label{fig:sensL}]{\includegraphics[width=0.5\textwidth]{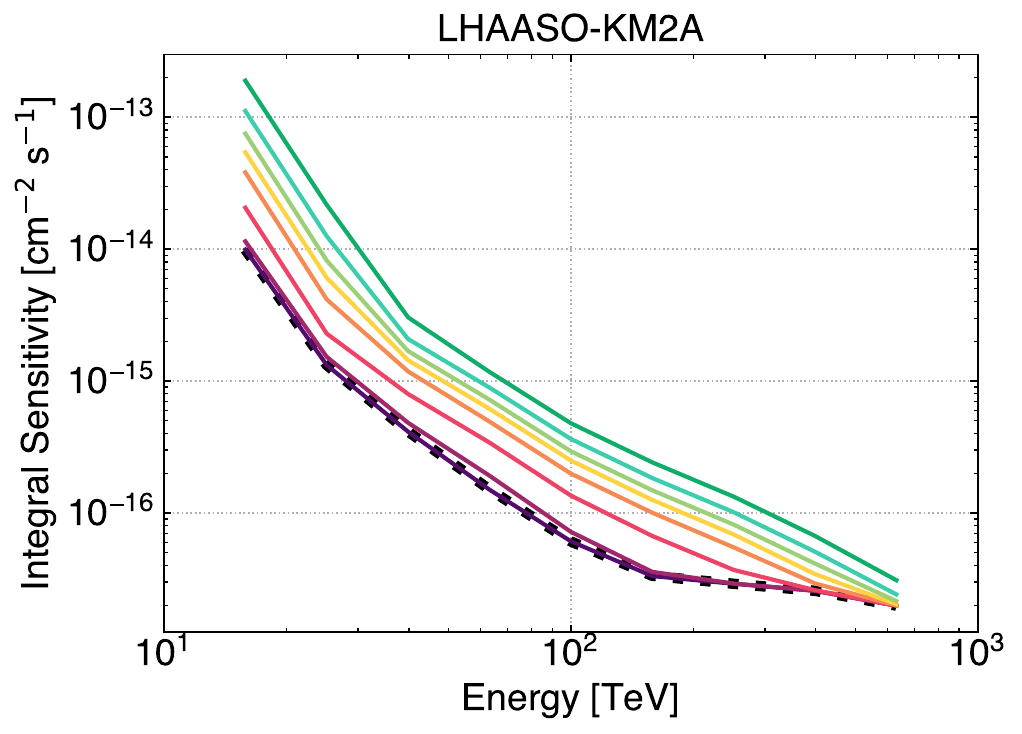}}
    \caption{Integral sensitivity for extended sources, as computed with the Crab-like PL spectrum defined in the text, i.e. with slope $\alpha=2.62$, by: (a) ASTRI, (b) CTA-N, (c) CTA-S, and (d) LHAASO. The exposure time adopted is 50 h for IACTs and 1 year for LHAASO. The source considered has a uniform disk-like morphology, whose angular radius is indicated in the legend, valid for all panels.}
    \label{fig:intsens}
\end{figure*}

\begin{figure*}
    \centering
    \subfigure[\label{fig:ratio_sens_diffindex}]{\includegraphics[width=0.45\textwidth]{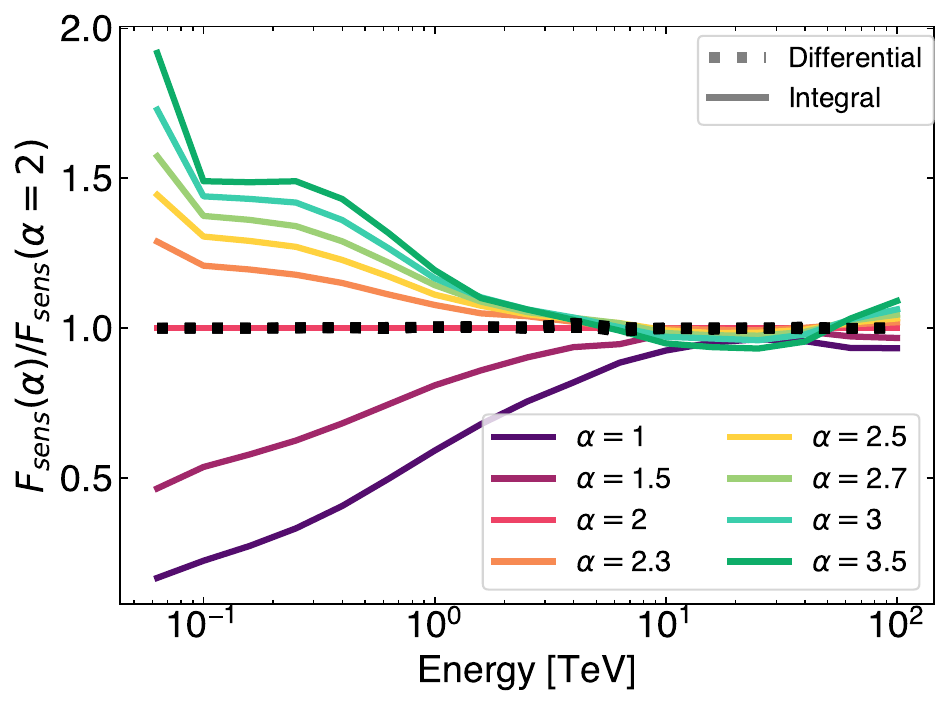}}
    \subfigure[\label{fig:ratio_sens_new}]{\includegraphics[width=0.45\textwidth]{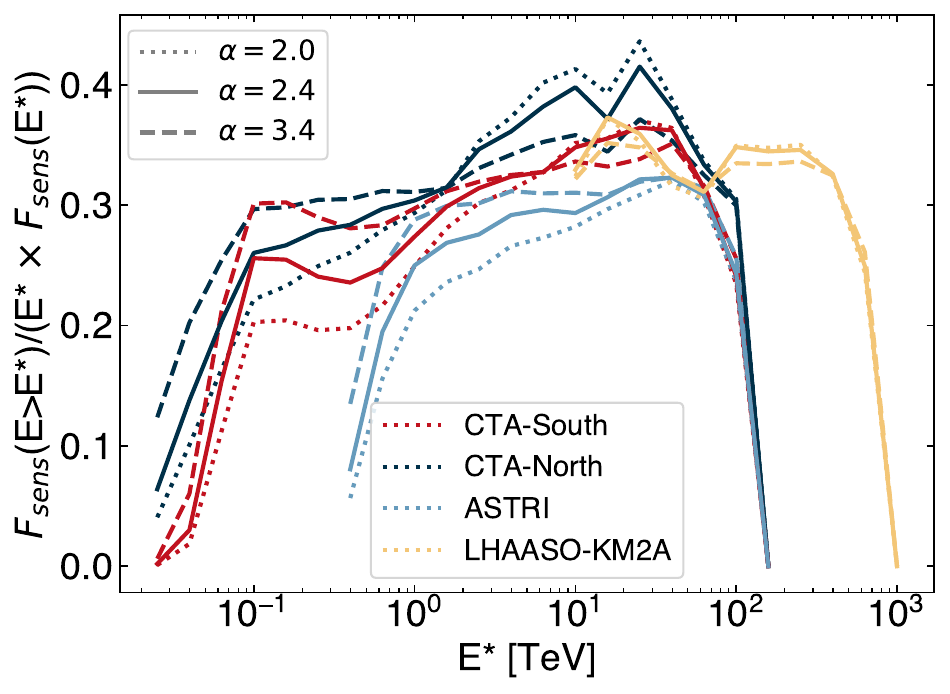}}
    \caption{\emph{Left:} Ratio between CTA-S integral (solid) or differential (dotted) sensitivities computed for different spectral slopes $\alpha$ (given in the legend) of a point-like source, and the corresponding one evaluated for $\alpha=2.0$. All the curves for differential sensitivities overlap, as expected. \emph{Right:} Ratio between point-like integral sensitivity (calculated for the spectral slopes indicated in the legend) and point-like differential sensitivity (multiplied by energy $E^*$) at various energy bins centred in $E^*$, for ASTRI, CTA and LHAASO-KM2A.}
    \label{fig:ratio_id}
\end{figure*}

An often adopted approach in gamma-ray astronomy is to compare integral source fluxes and sensitivities, rather than differential ones. This technique is chosen when dealing with weak sources, falling off the stringent bin-per-bin criteria set in Sec.~\ref{sec:sensitivity}. The computation of integral sensitivity, in fact, requires that the very same criteria are satisfied above a given energy threshold, $E^*$, namely that a certain number of events and an overall significance against background are obtained above $E^*$. The integral method should, however, be used with caution because of three main reasons: i) a comparative analysis of instruments that perform their best in different energy ranges might be misleading; ii) its results depend on the spectral slope of the target source, particularly at low energy as shown later, implying that a rigorous comparison among source flux and sensitivity can only be performed when the two refer to the very same spectral slope; and iii) as in the integration one loses the information about the spectral features beyond $E^*$, such as a cut-off, integral sensitivities might be overestimating the expected detection performance. In particular, this might be the case for LHAASO first catalogue sources, for which no information is currently available concerning the maximum photon energy detected for each of them. For these reasons, we adopted the differential approach as a benchmark. \\
Still, we find it worth to also discuss the results of the integral approach, as applied to extended sources, in order to understand its impact on source detectability. Being the criteria less stringent, a sensitivity improvement can in fact be expected, however the comparison with integrated fluxes yields a non trivial outcome, as we are going to discuss. Integral sensitivities for extended sources are provided in Fig.~\ref{fig:intsens}, as computed with the Crab-like spectrum defined in the text. \\
The impact of the chosen spectral slope, $\alpha$, both in the differential and in the integral sensitivity, $F_{\rm sens}(\alpha)$, is shown in the left panel of Fig.~\ref{fig:ratio_id} in terms of the ratio between $F_{\rm sens}(\alpha)$ calculated for CTA-South, and  $F_{\rm sens}(\alpha=2)$. Similar trends applies also to the other detectors. 
As expected, differential sensitivities are not affected by the slope of the spectrum, while integral sensitivities are very sensitive to this choice, and the major discrepancies emerge below $\sim$ 1 TeV. 
This result highlights that the differential approach is more preferable when discussing the detection prospects of sources with variegated spectral shapes, as it is the case of catalogues. On the other hand, if one focuses on the highest energies ($\gtrsim 10$ TeV), the dependence on the slope is almost negligible.

A proper comparison among differential and integral sensitivities can be achieved by multiplying the former to the central energy of the bin $E^*$ . The ratio of  integral sensitivity, $F_{\rm sens}(E>E^*,\alpha)$, to  $E^*\cdot F_{\rm sens}(E^*,\alpha)$  is shown in the right panel of Fig.~\ref{fig:ratio_id}, for different assumptions of source spectral slopes and for all considered detectors. The discussed slopes in this case coincide with the mean values of the distribution of slopes found for KM2A and HGPS sources, respectively $\alpha_{\rm KM2A} \simeq 3.4$ and $\alpha_{\rm HGPS} \simeq 2.4$. We also consider the $\alpha=2$ case for comparison.
As it can be seen, the usage of integral sensitivities can actually provide an improvement compared to the differential sensitivities at VHEs by an amount that depends on both energy and spectral slope of the source. Around 50 TeV, which is the reference value for following-up KM2A sources, the improvement is assessed around 30-35\% for IACTs. At lower energies, namely around 1 TeV, the improvement is slightly larger, up to factor 5, for the hardest considered spectra. However one should note that, as the integral sensitivity decreases, also the integral source flux gets reduced depending on its slope. In fact, the integral flux of a power-law function e.g. $F(E)= N_0(E/E_0)^{-\alpha}$ above a certain energy $E^*$ is 
$$F(>E^*)\approx \frac{N_0}{E^{-\alpha}_0} \frac{{(E^*)}^{1-\alpha}}{(\alpha-1)} = \frac{E^*~F(E^*)}{(\alpha-1)}$$ 
meaning that the source integral flux $F(>E^*)$ will be $\alpha-1$ times smaller than its differential energy flux $E^*\cdot~F(E^*)$, for any $\alpha > 1$.
In particular, for the LHAASO-KM2A sources, this factor amounts to $1/(\alpha_{\rm KM2A}-1)\sim 0.4$, while for HGPS sources it is $1/(\alpha_{\rm HGPS}-1)\sim 0.7$.
In both cases, as the flux reduction is more contained than the sensitivity gain, the integral approach actually provides a net gain in terms of detectability. Nonetheless, if compared at 50 TeV, the two effects are almost compensated for such that we do not envision an extension of the sample of KM2A sources that could be detected by IACTs, with respect to our results from Sec.~\ref{sec:catalog} with the differential approach. Conversely, at 3 TeV, the larger sensitivity gain from the integral approach is not expected to enlarge the sample of detectable sources, as most of them were already shown to be in the reach of IACTs with the more stringent differential approach, rather to reduce the exposure time required.

\section{IACT worsening of IRFs with offset}
\label{sec:appB}
\begin{figure*}
    \centering
    \includegraphics[width=0.9\linewidth]{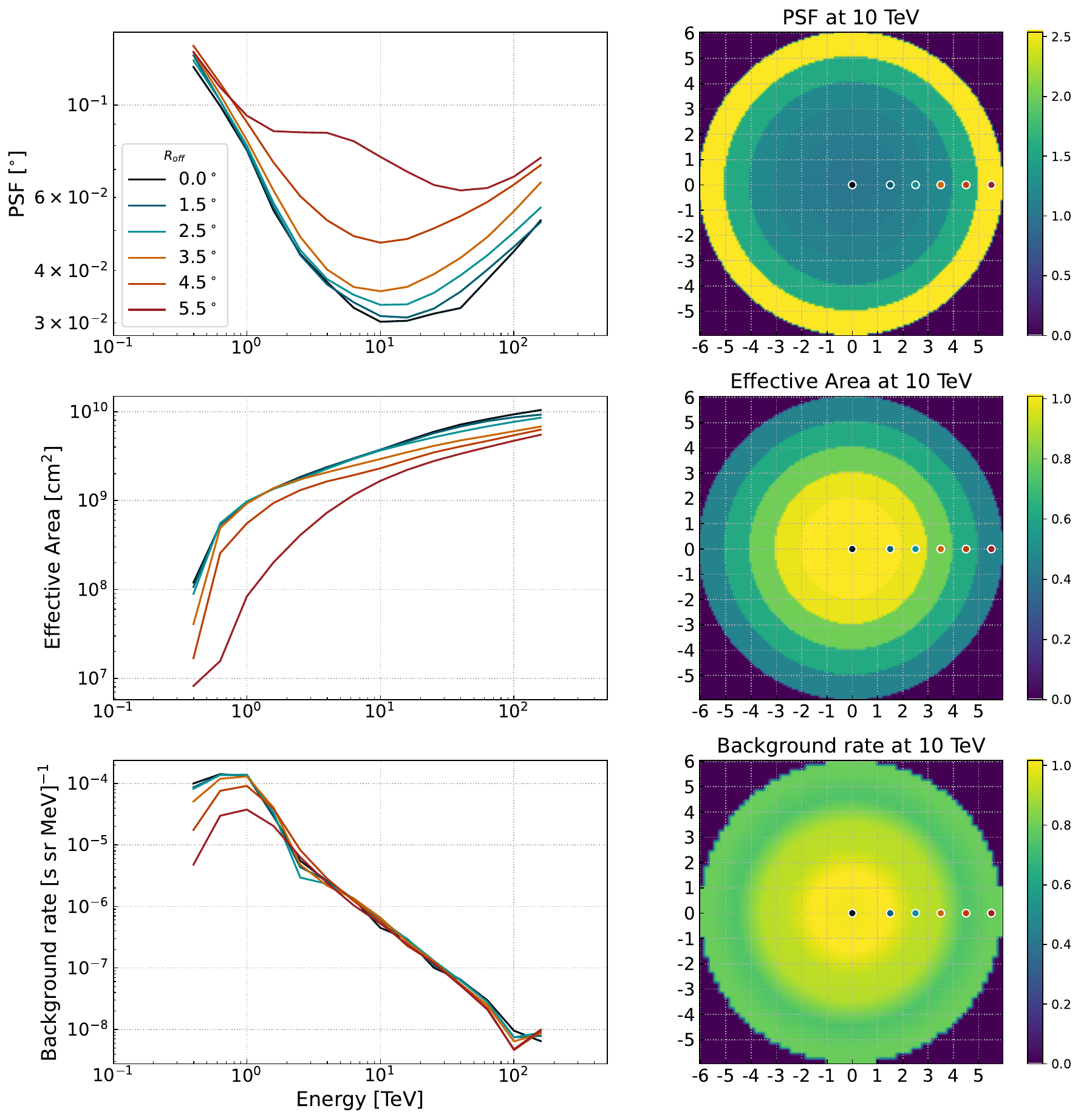}
    \caption{ASTRI instrument response functions (angular resolution, effective area, and background rate) at different offset angles from the centre of the FoV. In the left side, the IFRs are shown as a function of energy; in the right side, the degradation of the IRFs across the FoV is shown at a fixed energy of 10 TeV, where the responses are normalised to the value at the centre of the field. The legend applies to all panels.}
    \label{fig:astri_offset_IRFs}
\end{figure*}

\begin{figure*}
    \centering
    \includegraphics[width=0.9\linewidth]{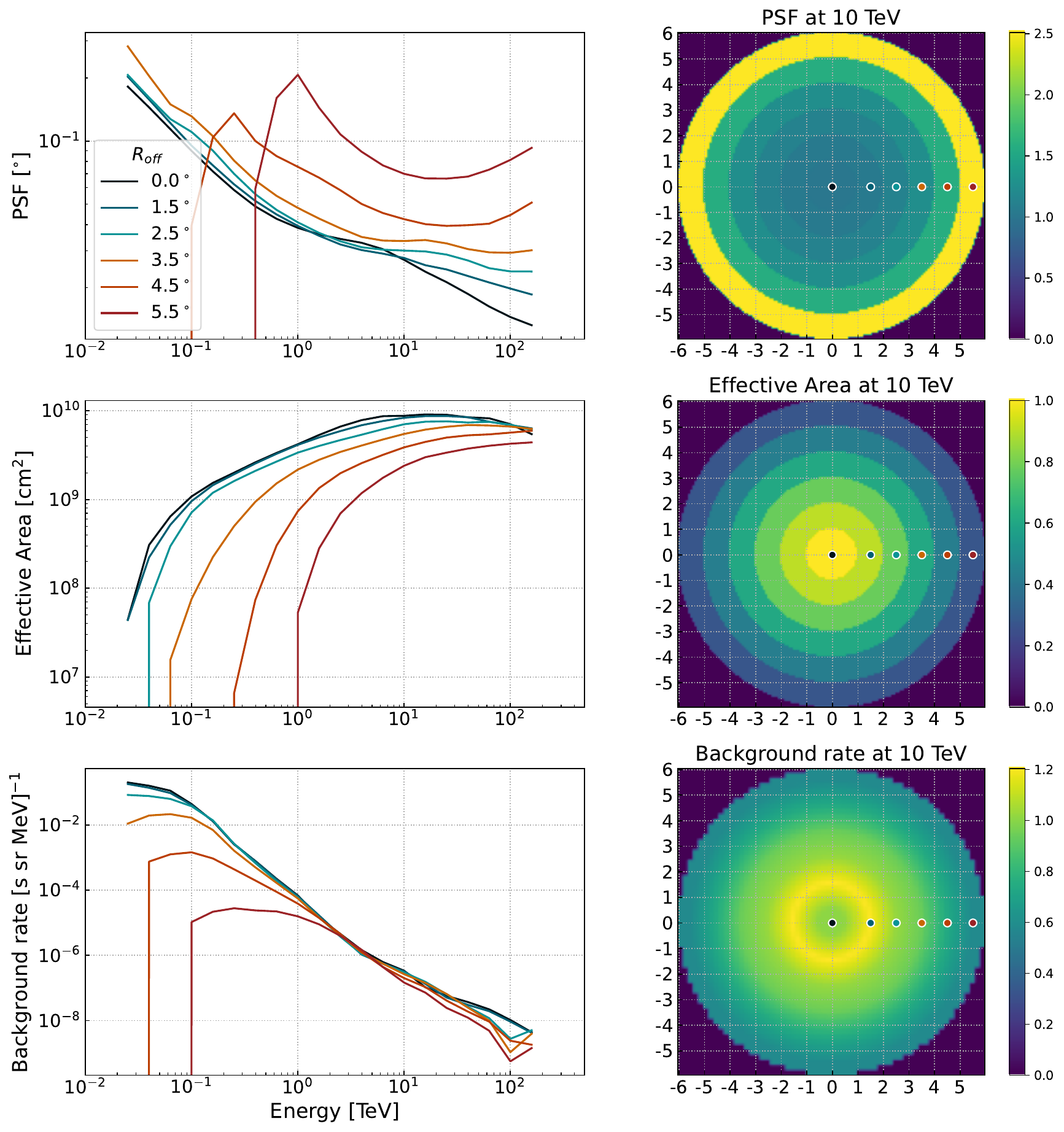}
    \caption{Same as in Fig.~\ref{fig:astri_offset_IRFs}, but for CTA-N IRFs. }
    \label{fig:ctan_offset_IRFs}
\end{figure*}

\begin{figure*}
    \centering
    \includegraphics[width=0.9\linewidth]{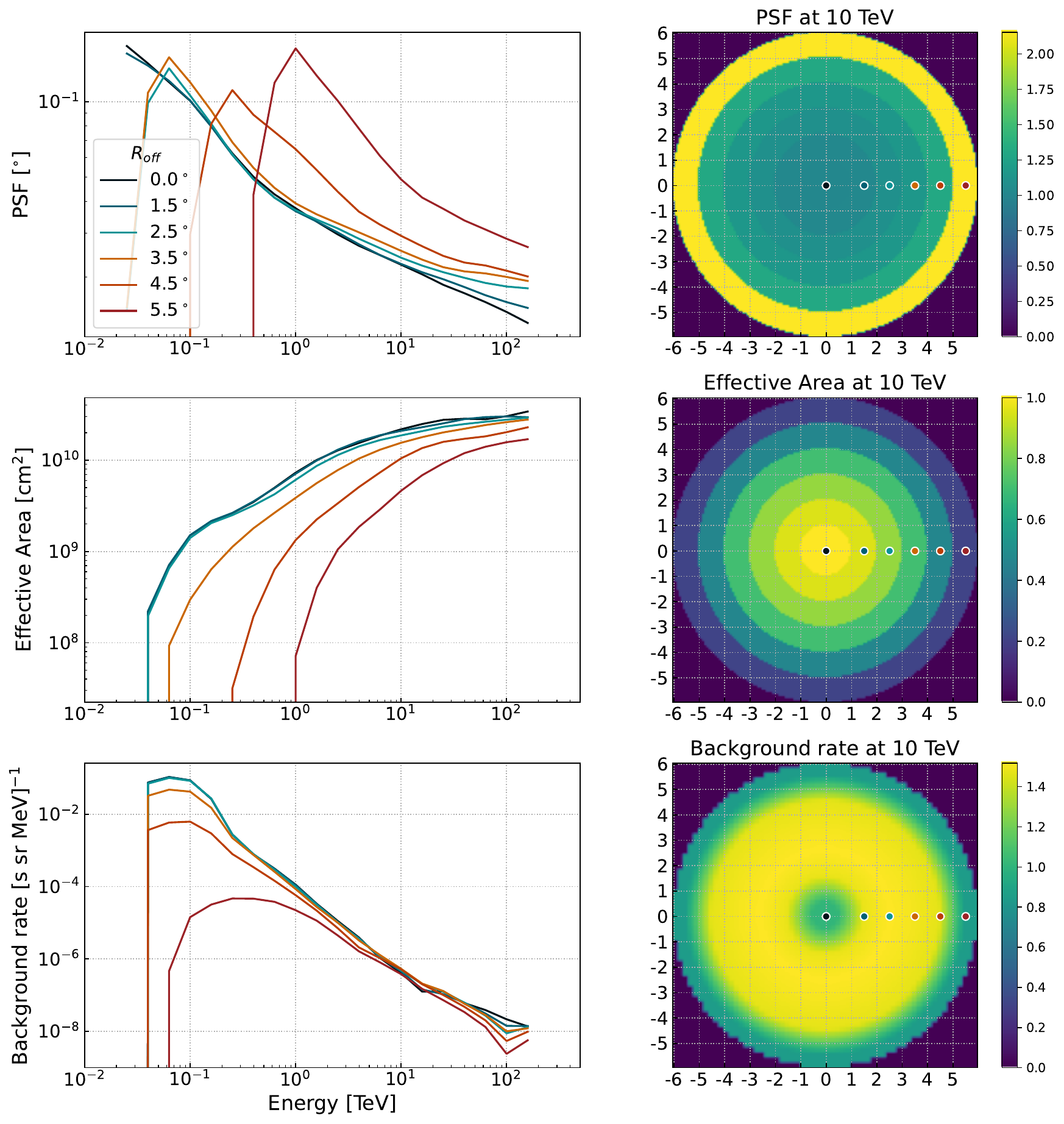}
    \caption{Same as in Fig.~\ref{fig:astri_offset_IRFs}, but for CTA-S IRFs. 
    }
    \label{fig:ctas_offset_IRFs}
\end{figure*}

\begin{figure*}[!ht]
    \centering
    \subfigure[]{\includegraphics[width= 0.49 \linewidth]{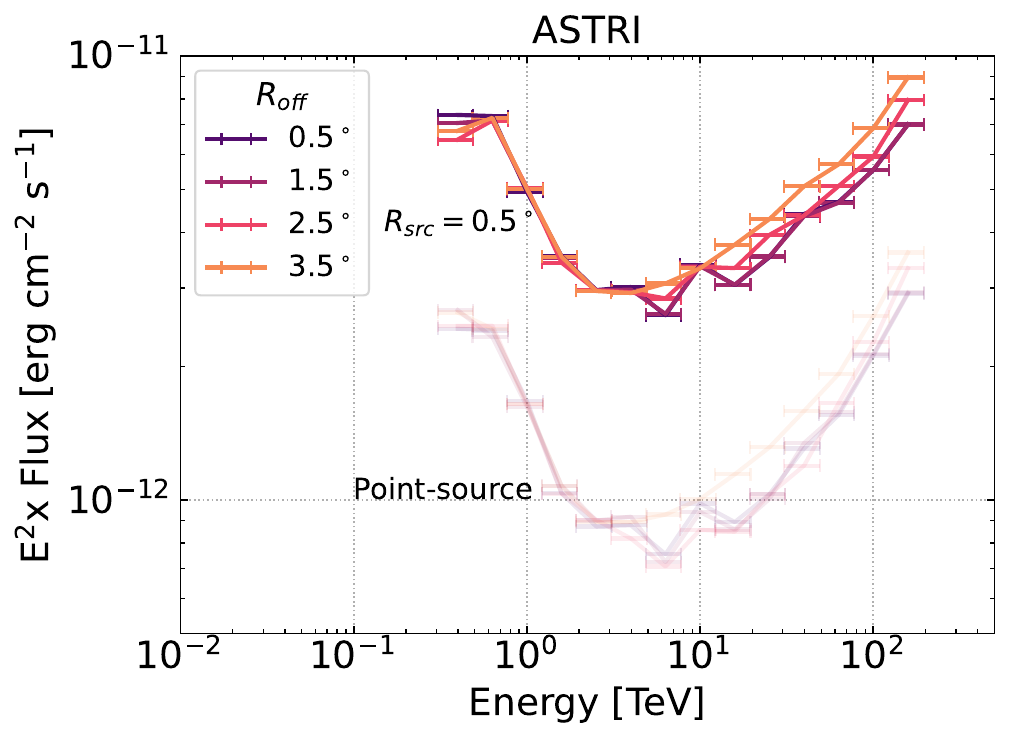}}  
    \subfigure[]{\includegraphics[width= 0.49 \linewidth]{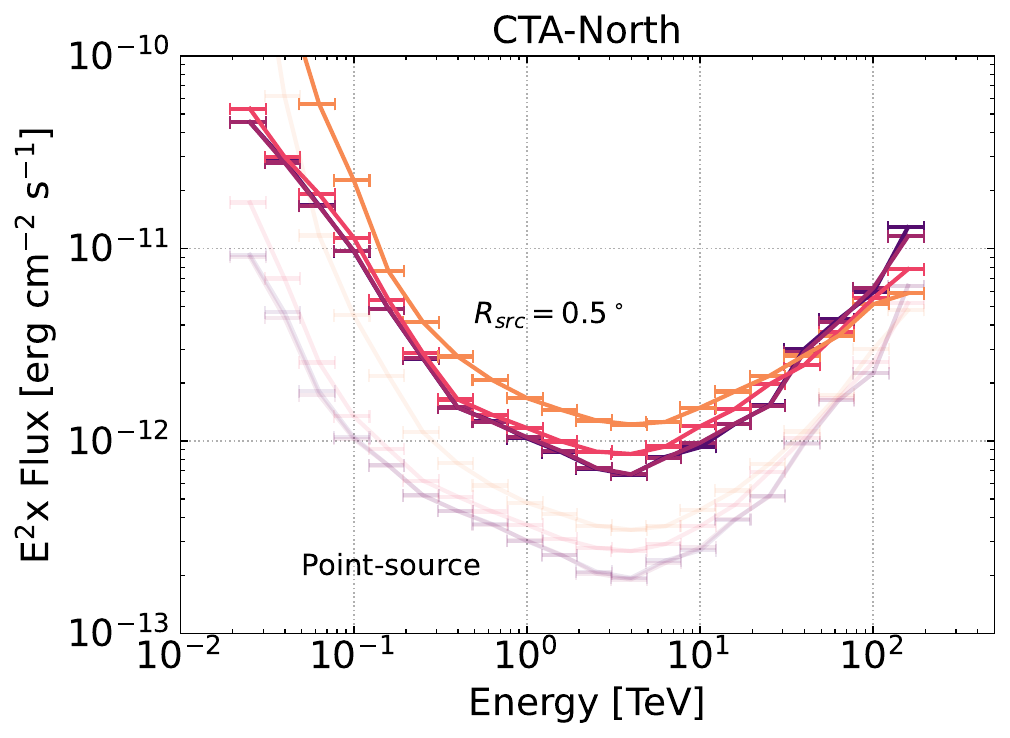}}
    \subfigure[]{\includegraphics[width= 0.49 \linewidth]{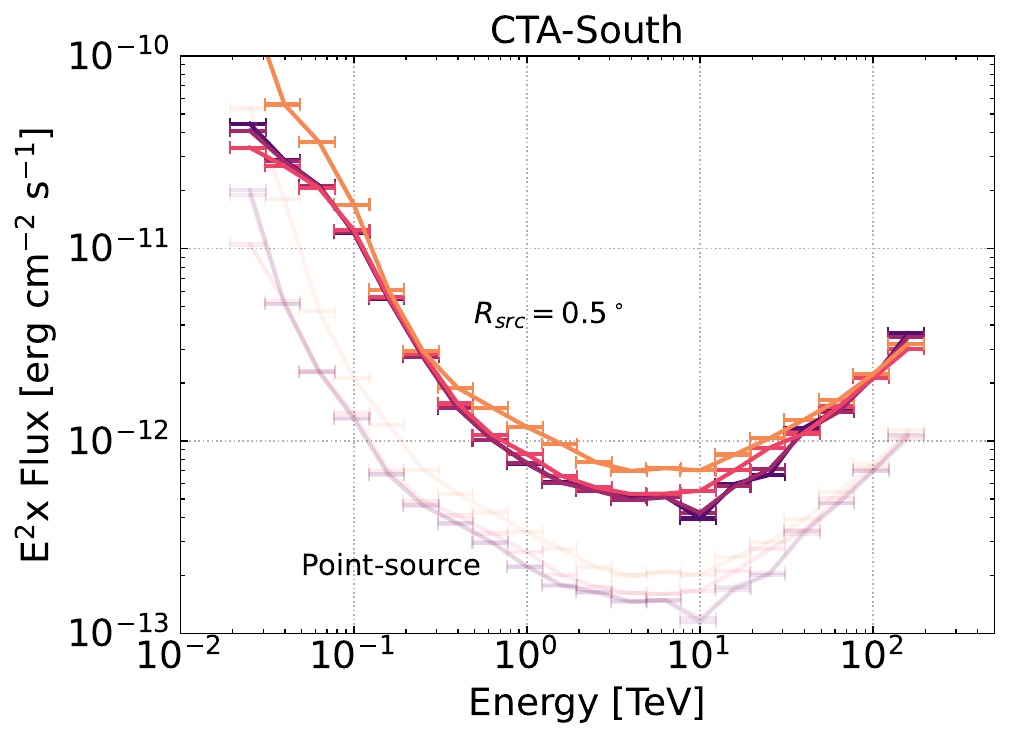}}
    \caption{Differential sensitivities of (a) ASTRI, (b) CTA-N, and (c) CTA-S to point-like sources (shaded) and extended ($R_{\rm src}=0.5^\circ$, solid) displaced with respect to the telescope axis by an offset angle $R_{\rm off}$, as indicated in the legend, which applies to all panels.}    
    \label{fig:offset_sens}
\end{figure*}

In the case of IACTs, extended source sensitivity is further degraded with respect to that presented in Sec.~\ref{sec:sensitivity} if observations are carried out   off-axis. Reduced performance are expected both in angular resolution and effective area. A preliminary estimation of the differential sensitivity degradation for extended source observations was performed in \citet{ambrogi2018} for the previous configuration of CTA-South: a limited worsening was obtained for sources with radial extension below 2~deg (within a factor 2 for energies above 50~GeV). Nowadays, public IRFs are available including off-axis responses, such that more detailed evaluations can be performed. In Figs.~\ref{fig:astri_offset_IRFs}, \ref{fig:ctan_offset_IRFs}, and \ref{fig:ctas_offset_IRFs} we show respectively ASTRI, CTA-North, and CTA-South performances as a function of the offset angle $R_{\rm off}$ of the observation with respect to the pointing direction of the telescope (fixed to zenith of 20~deg). For ASTRI, the impact of off-axis observations will mainly consist into a reduced effective area and enlarged angular resolution. For CTA, in addition, an increase in the energy threshold is expected. \\
The differential sensitivities are consequently affected by off-axis observations. We performed analogous calculations to those presented in Sec.~\ref{sec:sensitivity}, this time using as input the off-axis performance. We show the results in Fig.~\ref{fig:offset_sens} for case off-axis observations of both point-like sources and $R_{\rm src}=0.5^\circ$ extended.

\newpage
\newpage

\section{IACT sensitivity for high zenith angle observations}
\label{sec:appC}
\begin{figure}
    \centering
    \includegraphics[width=1\linewidth]{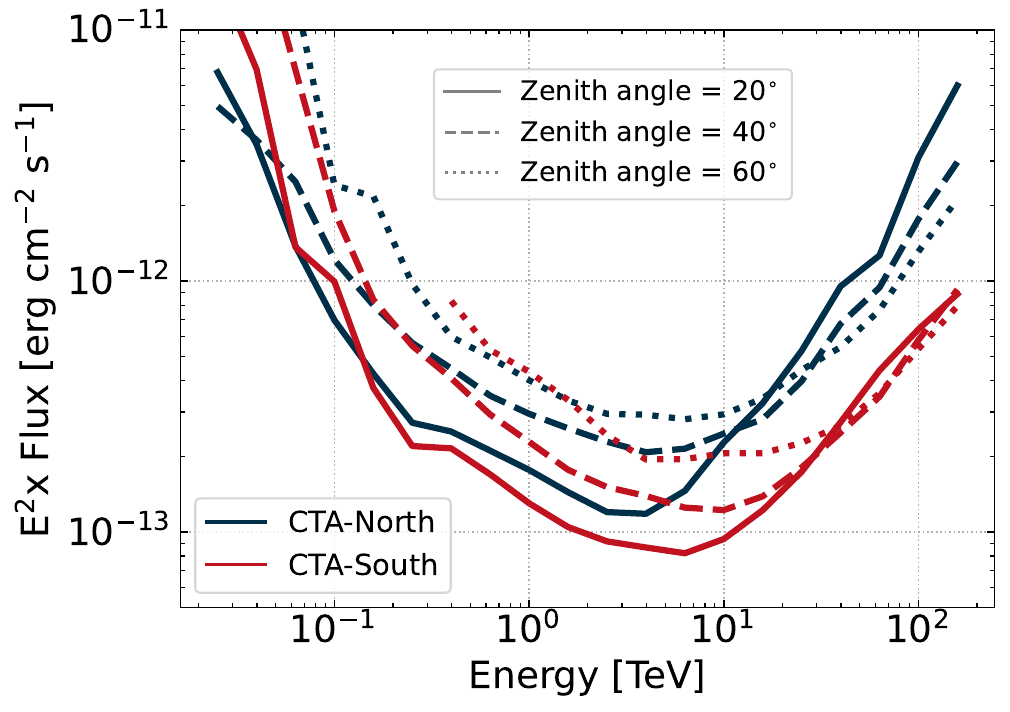}
    \caption{Differential sensitivity to point-like source observations for CTA-N and CTA-S at different  zenith angles.}
    \label{fig:sens_zenith}
\end{figure}

IACT observations at high-zenith angles are characterised by a larger effective area at high energies at the expenses of an increased energy threshold. In fact, high-energy quasi-horizontal showers can profit of a better visibility, thanks to the larger portion of the atmosphere where they develop. On the other hand, low-energy high-zenith showers are less efficiently detected as their first interaction vertex occurs further away from the telescopes. As a result, the differential sensitivity of IACTs are affected by observations of sources at different zenith angles during the acquisition. We quantify the impact of this effect by computing the expected sensitivity for point-like observations of the CTA-N and CTA-S observatories, as shown in Fig.~\ref{fig:sens_zenith}. As visible, an improvement of a factor 2 is expected above a few tens of TeV for CTA. We expect similar considerations to hold for ASTRI as well, however we cannot proceed to a proper evaluation of it since IRFs at high-zenith angle observations are not yet publicly available. 

\section{Point-like differential sensitivities: Comparison with official results and investigation of CTA subarrays}
\label{sec:appD}
\begin{figure}[ht!]
    \centering
    \includegraphics[width=1\linewidth]{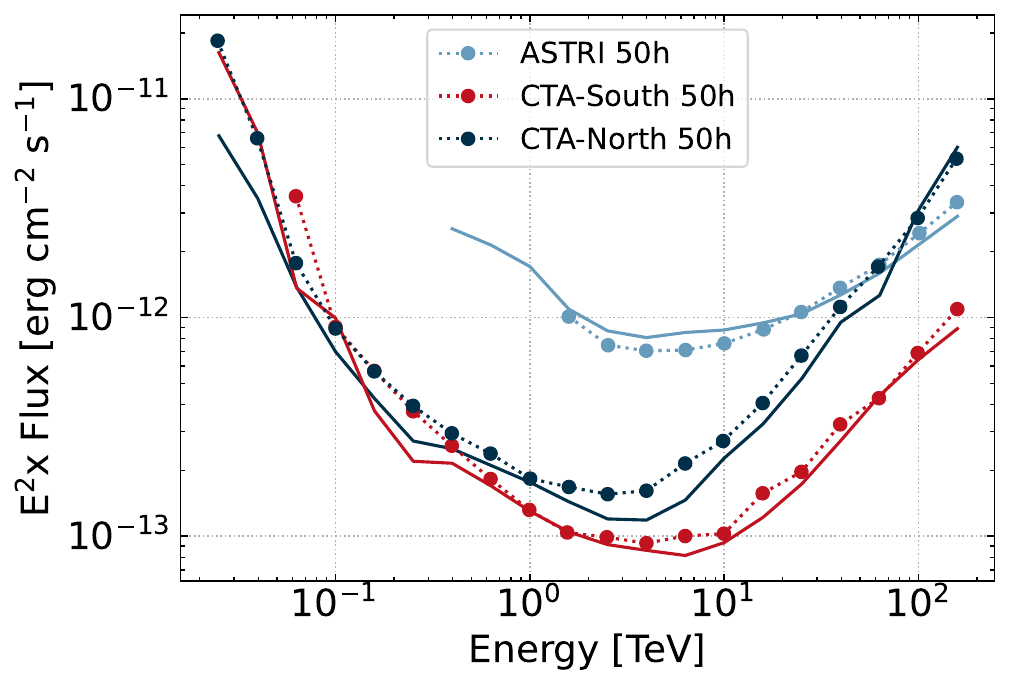}
    \caption{Point source differential sensitivity calculated in this work (solid lines) compared to the point source sensitivities released by the CTAO \citep{cta2022IRFs} and ASTRI Collaborations \citep{Vercellone2023Astri} (dotted lines). All sensitivities are calculated for an exposure 50 hours, at 20$^\circ$ zenith angle.}
    \label{fig:sens_ps}
\end{figure}

\begin{figure}[ht!]
    \centering
    \includegraphics[width=1\linewidth]{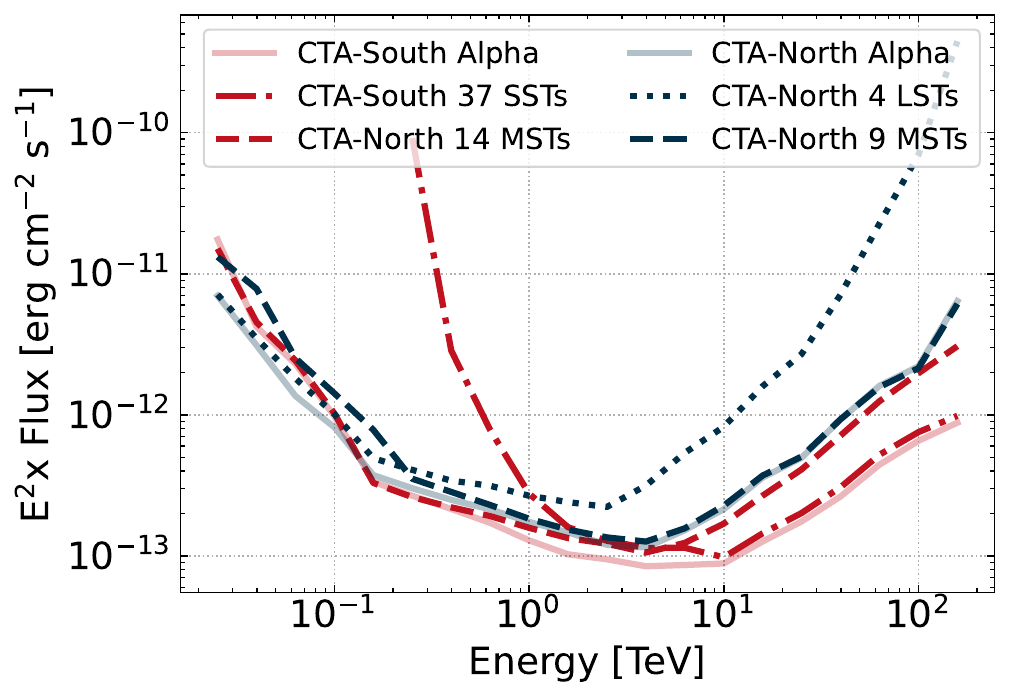}
    \caption{Differential sensitivity of the CTA-Alpha array compared to its sub-array configurations, as indicated in the figure legend.}
    \label{fig:sub_sens}
\end{figure}

A cross-check of the computation methods here presented for differential sensitivities can be obtained by comparing our on-axis point-like results with the official ones released by the Collaborations \citep{cta2022IRFs,2022icrc.confE.884L,Vercellone2023Astri}, as in Fig.~\ref{fig:sens_ps}. The good agreement emerging for all the observatories testifies the solidity of the methods here developed and hence its applicability to extended source studies. \\
For the interested reader, we further provide in Fig.~\ref{fig:sub_sens} the calculated point-like sensitivities of the different CTA-Alpha sub-arrays (SSTs, MSTs, and LSTs), as computed from the official Consortium release IRFs \citep{cta2022IRFs}, showing the role of each sub-array across the whole energy range probed by CTA.

\end{appendix}

\end{document}